\documentclass[12pt]{article}
\usepackage{latexsym}
\usepackage{cite,epsfig,amssymb,euscript,slashed}
\usepackage{amsmath}
\usepackage{array,calc,epsfig}
\usepackage{bbm}
\usepackage{fancybox}

\def\be{\begin{equation}}
\def\ee{\end{equation}}
\def\bseq{\begin{subequations}}
\def\eseq{\end{subequations}}

\def\bea{\begin{eqnarray}}
\def\eea{\end{eqnarray}}

\def\bseq{\begin{subequations}}
\def\eseq{\end{subequations}}

\numberwithin{equation}{section} %%
\usepackage[]{graphicx}

\def\d {{\rm d}}

\def\cald         {{\cal D}}
\def\cale         {{\cal E}}
\def\calf         {{\cal F}}
\def\calg         {{\cal G}}

\def\calk         {{\cal K}}
\def\call         {{\cal L}}
\def\calm         {{\cal M}}
\def\caln         {{\cal N}}

\def\calr         {{\cal R}}

\def\calw         {{\cal W}}
\def\calz         {{\cal Z}}

\def\del          {\partial}

\def\ii           {{\rm i}}

\def\Re           {{\rm Re\hskip0.1em}}
\def\Im           {{\rm Im\hskip0.1em}}

\def\sqr#1#2{{\vcenter{\vbox{\hrule height.#2pt
 \hbox{\vrule width.#2pt height#1pt \kern#1pt \vrule width.#2pt}\hrule
 height.#2pt}}}}

\def\d{\text{d}}

\def\slashchar#1{\setbox0=\hbox{$#1$}           % set a box for #1
\dimen0=\wd0                                 % and get its size
\setbox1=\hbox{/} \dimen1=\wd1               % get siste of /
\ifdim\dimen0>\dimen1                        % #1 is bigger
\rlap{\hbox to \dimen0{\hfil/\hfil}}      % so center / in box
#1                                        % and print #1
\else                                        % / is bigger
\rlap{\hbox to \dimen1{\hfil$#1$\hfil}}   % so center #1
/                                         % and print /
\fi}

\begin{document}
\font\cmss=cmss10 \font\cmsss=cmss10 at 7pt

\title{\vspace{-2.0cm}\begin{flushright}{\scriptsize DFPD-2017/TH/10}
%\\  \scriptsize  preprint2}
\end{flushright}
\hfill\\
\hfill\\
Three-forms in Supergravity 
\\[0.1cm] 
and Flux Compactifications
\\[0.5cm] }

\author{Fotis Farakos, Stefano Lanza, Luca Martucci and Dmitri Sorokin}

\date{}

\maketitle

\vspace{-1.5cm}

\begin{center}

\vspace{0.5cm}
\textit{\small  Dipartimento di Fisica e Astronomia ``Galileo Galilei",  Universit\`a degli Studi di Padova \\ 
\& I.N.F.N. Sezione di Padova, Via F. Marzolo 8, 35131 Padova, Italy}
\end{center}

\vspace{5pt}

\abstract{
\noindent
We present a duality procedure that relates conventional four-dimensional matter-coupled $\caln=1$  supergravities to dual formulations in which auxiliary fields are replaced by field-strengths of gauge three-forms. The duality promotes specific coupling constants appearing in the superpotential to  vacuum expectation values of the field-strengths.    We then apply this general duality to  type IIA string compactifications on Calabi-Yau orientifolds with RR fluxes. This gives a new supersymmetric formulation of the corresponding effective four-dimensional theories which includes gauge three-forms. }

%\noindent {\em Possible comment ..........................................................................................................................................}

\thispagestyle{empty}

%\vfill
%\vskip 5.mm
%\hrule width 5.cm
%\vskip 2.mm
%{\scriptsize
%\noindent e-mails: {\tt massimo.bianchi@roma2.infn.it, andres.collinucci@physik.uni-muenchen.de, luca.martucci@roma2.infn.it
%}}

\newpage

\setcounter{footnote}{0}

\tableofcontents

\newpage

\section{Introduction}

The physical role of gauge three-forms in  four-dimensional field theories has been under study for several decades. For instance, constant four-form fluxes of these fields may effect the value of the cosmological constant directly or via couplings of the three-forms to membranes (see e.g.\ \cite{Duff:1980qv,Aurilia:1980xj,Hawking:1984hk,Brown:1987dd,Brown:1988kg,Duff:1989ah,Duncan:1989ug,Ovrut:1997ur,Bousso:2000xa,Feng:2000if,Wu:2007ht,Bandos:2010yy,Bandos:2011fw,Bandos:2012gz,Farakos:2016hly}). 
A possible role of  three-forms in the solution of the strong CP problem was discussed e.g. in  \cite{Dvali:2005an,Dvali:2004tma,Dvali:2005zk,Dvali:2013cpa,Dvali:2016uhn,Dvali:2016eay} and in inflationary models in \cite{Kaloper:2008fb,Kaloper:2011jz,Marchesano:2014mla,Bielleman:2015ina,Dudas:2014pva,Valenzuela:2016yny}. 
In the context of four-dimensional global and local supersymmetric theories, three-form gauge fields can be  naturally incorporated as auxiliary fields of supermultiplets, as  e.g.\ in \cite{Stelle:1978ye,Ogievetsky:1978mt,Ogievetsky:1980qp,Gates:1980ay,Gates:1980az,Buchbinder:1988tj,Binetruy:1996xw,Ovrut:1997ur,Kuzenko:2005wh,Nishino:2009zz,Duff:2010vy,Bandos:2010yy,Bandos:2011fw,Bandos:2012gz,Groh:2012tf,Farakos:2016hly,Bandos:2016xyu,Aoki:2016rfz,Buchbinder:2017vnb}.

Furthermore, effective field theories with gauge three-forms can find a natural application  in the context of string compactifications \cite{Grimm:2011dx,Kerstan:2011dy,Bielleman:2015ina,Carta:2016ynn}. In particular, the effective four-dimensional theories describing flux compactifications of type IIA and IIB string theories should allow for a supersymmetric formulation including gauge three-forms, whose field-strengths are dual to the fluxes threading the internal compactified  space.  In \cite{Bielleman:2015ina} it was suggested that three-forms coming from the dimensional reduction of type II supergravities could be associated with auxiliary fields of chiral and gravity multiplets.   
However, this idea does not seem to be realizable within any of the  four-dimensional supersymmetric models constructed so far.

This problem motivated us to revisit the role of gauge three-forms in four-dimensional rigid and local supersymmetry, focusing on the minimal $\caln=1$ case and looking for supergravity-matter models in which the results of \cite{Bielleman:2015ina} could fit. More specifically, we will address the following general question, suggested by the somewhat universal structure of the four-dimensional effective theories describing string flux compactifications. Consider a supersymmetric theory with a set of  chiral superfields $\Phi^A$ and a superpotential of the form
\be\label{sup0}
W=e_A\Phi^A+m^A\calg_{AB}(\Phi)\Phi^B+\hat W(\Phi) \, , 
\ee
where $e_A$ and $m^A$ are real constants, $\hat W(\Phi)$ and  $\calg_{AB}(\Phi)$ are arbitrary holomorphic functions which, even if not explicitly indicated, can possibly depend on additional chiral superfields. The question is then: does there exist an alternative supersymmetric formulation of the effective theory with a set of pairs of gauge three-forms $(A^A_3,\tilde A_{3 A})$ in which the coupling constants $e_A$ and $m^B$ are promoted to  vacuum expectation values of the field-strengths $F^A_4=\d A^A_3$ and $\tilde F_{4A}=\d \tilde A_{3A}$?
Note that this procedure is a certain kind of duality transformation that   trades coupling constants for  gauge three-forms,  which do not carry propagating degrees of freedom in four dimensions.

In this paper we will provide a positive answer to this question. The new formulation will be obtained by a  supersymmetric duality transformation, which modifies the structure of the chiral multiplets $\Phi^A$, substituting their scalar complex auxiliary fields $F^A$ or just the real parts thereof  with a  combination of the field-strengths  $F^A_4$ and $\tilde F_{4A}$.  Furthermore, this procedure naturally generalizes to the locally supersymmetric case when one of the scalar superfields $\Phi^A$  (e.g. $\Phi^0$) is considered to be the compensator of the super-Weyl-invariant formulation of supergravity. After gauge-fixing the super-Weyl symmetry, the duality transformation involves also the auxiliary field of the old minimal supergravity multiplet.

Before arriving at the detailed discussion of the general dualization procedure outlined above, we will first consider the simpler subcases in which  $\calg_{AB}$ is constant. In these subcases, our dualization  explicitly relates the three 
known types of chiral multiplets: the conventional one with the complex scalar as the auxiliary field, the {\em single three-form multiplet} in which the complex auxiliary field is a sum of a real scalar and the Hodge dual of the field-strength of a real gauge three-form, and the {\em double three-form multiplet} in which the auxiliary field is the field-strength of a complex gauge three-form. In particular, the single three-form multiplets arise when the matrix $\Im\calg_{AB}$ is degenerate, as for instance in the extreme case $\Im\calg_{AB}\equiv 0$. 

In the case of constant $\calg_{AB}$  the relation between the conventional chiral  and the dual three-form multiplet is linear. This is no longer true for a general $\calg_{AB}(\Phi)$ in which case the duality relation is non-linear and might not allow for a general explicit superfield solution. However it turns out to be  tractable if we assume that $\calg_{AB}(\Phi)$ is identified with the second derivative of a homogeneous ``prepotential" $\calg(\Phi)$ of degree two. In fact, this is what happens in string flux compactifications.  

In the course of the study of the dual formulations with  three-form multiplets we will encounter a subtlety regarding the presence of  boundary terms in the Lagrangian. The necessity to take into account appropriate boundary terms  in the theories with gauge three-forms, either supersymmetric or not,  is well known (see e.g. \cite{Brown:1987dd,Duff:1989ah,Duncan:1989ug,Groh:2012tf}).  As we will show, our dualization procedure automatically produces the correct boundary terms, which then do not need to be introduced by hand. 
 
As a concrete non-trivial example, we will perform the duality transformation of the supersymmetric effective theory associated with type IIA orientifold  string compactifications on Calabi-Yau spaces with Ramond-Ramond (RR) fluxes. 
This effective theory has a superpotential of the form (\ref{sup0}) with $\calg_{AB}(\Phi)=\del_A\del_B\calg(\Phi)$ and $\calg(\Phi)$ being homogeneous of degree-two. In this superpotential the constants $e_A$ and $m^B$ are identified with the quanta of the internal RR fluxes threading the compactification space and $\Phi^A$ with a combination of the K\"ahler moduli and the super-Weyl compensator superfields. As we will see,  the field-strengths  $F^A_4$ and $\tilde F_{4A}$ produced by the duality procedure perfectly match the field-strengths obtained by direct dimensional reduction  of the IIA RR field-strengths   in \cite{Bielleman:2015ina}. For simplicity, we will work under the assumption that the internal NSNS flux vanishes, which allows us to ignore the tadpole cancellation condition. For more general type IIA, as well as type IIB flux compactifications, the tadpole condition must be appropriately taken into account. Furthermore, the dual formulation with gauge three-forms should allow for a natural incorporation of the open-string sector into the effective theory, as in \cite{Grimm:2011dx,Kerstan:2011dy,Carta:2016ynn}. We leave these interesting developments for the future.

The paper is organized as follows. In Section 2 we introduce the duality procedure in rigid supersymmetric theories. We first discuss simpler cases with constant $\calg_{AB}$, reviewing the structure of the corresponding known types of chiral three-form multiplets.  We then generalize the dualization procedure to a general $\calg_{AB}(\Phi)$, which leads to a non-linear duality relation.

In Section 3 we extend the duality procedure to supergravity. We first apply it to  pure old-minimal $\caln=1$ supergravity in its super-Weyl invariant formulation, producing the three-form formulations thereof. In particular,  this shows how the different formulations  are related to each other by duality transformations of the corresponding super-Weyl compensators. 
Then  we consider models with chiral multiplets coupled to supergravity and apply to them the non-linear duality transformation put forward in the rigid case. The duality  acts simultaneously on  matter superfields and the super-Weyl compensator. In the resulting dual formulation the auxiliary fields of the chiral and gravity multiplets are expressed in terms of the gauge three-forms and the scalar fields. 

In Section 4 we apply the duality transformation  to the effective four-dimensional theory associated with orientifold type IIA string compactifications with RR fluxes.  We also provide the explicit relation between  field-strengths of the four-dimensional theory and the ten-dimensional RR  fields. 

In Appendix \ref{B} we give the component content of the different four-dimensional $\caln=1$ superfields which are used in the main text.
In Appendix \ref{A} we show how the dualization procedure works for a simple bosonic field theory and then consider an instructive example which explains how the bosonic boundary terms can be obtained as components of a superspace defined Lagrangian.
Appendix \ref{C} contains useful expressions for the applications to  type IIA flux compactifications.

We mainly use notation and conventions of \cite{Wess:1992cp}.

\section{Three-form multiplets in supersymmetry}\label{rigid}

In this section we explain how the dualization procedure works in the case of rigid $\caln=1$ supersymmetric theories. In the simplest case of constant $\calg_{AB}$ in (\ref{sup0}), it will produce known variants of off-shell chiral multiplets, whose auxiliary fields are replaced by the field-strength of one or two gauge three-forms. We will refer to these chiral multiplets as {\em single} and {\em double} three-form multiplets, respectively. As we will see, in the case of generic $\calg_{AB}(\Phi)$, the dualization will provide a generalization of these off-shell three-form multiplets.

\subsection{Single three-form multiplets}\label{1-4}

Consider a rigid supersymmetric theory for a set of chiral superfields 
\be\label{Phi}
\Phi^A=\varphi^A+\sqrt{2}\theta\psi^A+\theta^2 F^A \, , 
\ee
with a superpotential of the form (\ref{sup0}) in the simplest case in which $\Im\calg_{AB}=0$. In such a case, since $\calg_{AB}$ is holomorphic, $\Re\calg_{AB}$ is necessarily constant and then the Lagrangian takes the form
\be\label{LPhir}
{\mathcal L}=\int \d^4 \theta K(\Phi,\bar\Phi)+\Big(\int \d^2\theta\big[ r_A\Phi^A+\hat W(\Phi)\big]+\text{c.c.}\Big)\,.
\ee
where $r_A\equiv e_A+m^B\Re\calg_{AB}$ are {\em real} constants.

To dualize the Lagrangian \eqref{LPhir},  we promote the constants $r_A$ to chiral superfields $X_A$ and introduce real scalar superfields $U^A$ as Lagrange multipliers. The  Lagrangian (\ref{Phi}) gets substituted by  
\be\label{LPhiXr}
\begin{aligned}
{\mathcal L}'=&\int \d^4\theta\,  K(\Phi,\bar\Phi)+\Big(\int \d^2\theta\, X_A\Phi^A+\text{c.c.}\Big)+\ii\int \d^4\theta  (X_A-\bar X_A)U^A\\
&+\Big(\int \d^2\theta\, \hat W(\Phi)+\text{c.c.}\Big)\,.
\end{aligned}
\ee
Integrating out $U^A$ by imposing its equations of motion one gets
\be\label{X+bX}
X_A-\bar X_A=0\,.
\ee
The chirality of $X_A$ ($\bar D_{\dot \alpha} X_A=0=D_\alpha \bar X_A$) then implies that $X_A=r_A$, with $r_A$ being real constants. Plugging this solution back into \eqref{LPhiXr} we get the initial Lagrangian \eqref{LPhir}.

To find the formulation of the theory in terms of three-form multiplets we vary \eqref{LPhiXr} with respect to  $X_A$ subject to the boundary conditions 
\begin{equation}\label{Bound_3F_Cond}
\delta X_A |_{\mathrm{bd}} = 0 \, , 
\end{equation}
which gives
\be\label{Phi=DDV}
\Phi^A=Y^A \, , 
\ee
with
\be\label{defY}
Y^A\equiv \frac \ii 4 \bar D^2 U^A\,. 
\ee
The superfields  $Y^A$ differ from ordinary chiral superfields only in their $\theta^2$-components 
\be\label{Y}
Y^A=y^A+\sqrt{2}\theta\chi^A+\theta^2({}^*\!F^A_4+\ii D^A)\,,
\ee
where $D^A$ are real auxiliary scalar fields and 
\begin{equation}
F^A_4=\d A^A_3 \, . 
\end{equation} 
Hence the real-part of the ordinary scalar auxiliary fields is substituted by the field-strengths $F^A_4$ of the gauge three-form $A^A_3$, which are part of the $U^A$ multiplets (see Appendix \ref{B}). 
The three-form fields appear only inside their field strengths because of the invariance of \eqref{defY} under the gauge transformations
\be\label{Ug}
U^A \quad \to \quad U^A+L^A \, ,
\ee
where $L^A$ are arbitrary real linear superfields  $D^2L^A=\bar D^2L^A=0$. This superspace gauge symmetry incorporates the bosonic gauge symmetry
\be
A_3^A\rightarrow A^A_3+\d\Lambda^A_2 \, , 
\ee
and mods out the redundant components of $U^A$ which do not survive the chiral projection $\frac \ii 4 \bar D^2$.  

We will refer to the chiral superfields $Y^A$ as single three-form multiplets. 
This kind of scalar multiplet was introduced in \cite{Gates:1980ay} and studied in detail in \cite{Groh:2012tf}. For instance, \cite{Groh:2012tf} studied the  relation of these multiplets with other multiplets, in particular, with conventional chiral multiplets.  The above simple duality argument  explicitly shows how the conventional  chiral multiplets and three-form multiplets  are related in a manifestly supersymmetric way.

To complete the dualization procedure, we should also take into account the equations of motion of $\Phi^A$ obtained from (\ref{LPhiXr}) with $\delta \Phi^A$ subject to the boundary condition $\delta\Phi^A |_{\mathrm{bd}} = 0$. These give the expression for $X_A$ in terms of $Y^A$
\be\label{Bound_3FAux_1}
X_A = \frac14 \bar{D}^2K_A(Y)-\hat W_A (Y) \, , 
\ee
where $K_A\equiv \del_A K$ and $W_A\equiv\del_A W$.

Upon plugging \eqref{Phi=DDV} and \eqref{Bound_3FAux_1} into \eqref{LPhiXr} we get the dual Lagrangian describing the dynamics of the superfields $Y^A$
\be\label{LY}
\hat{\mathcal L}=\int \d^4 \theta\, K(Y,\bar Y)+\Big(\int\d^2\theta\, \hat W(Y)+\text{c.c.}\Big)\,+{\mathcal L}_{\rm bd},
\ee
where
\be\label{bt0}
{\mathcal L}_{\rm bd}= 
\ii\int\d^2\theta \left(\int\d^2\bar\theta+\frac14\bar D^2\right)\left(\left(\frac14 \bar D^2 K_A - \hat{W}_A \right)U^A\right) 
+\text{c.c.}  
\, , 
\ee
is a total derivative and hence  a boundary term. Notice that in (\ref{LY}) there is no $r_A\Phi^A$ term in the superpotential. Furthermore, in general the boundary term (\ref{bt0})  gives a non-vanishing contribution to the Lagrangian and hence cannot be neglected.\footnote{Note that the Lagrangians (\ref{LPhiXr}) and (\ref{LY}) are gauge invariant under \eqref{Ug} provided  $X_A$ satisfy the boundary conditions  $X_A|_{\rm bd}=r_A$, where $r_A$ are (at least, classically) arbitrary real constants which characterize the asymptotic vacuum  of the theory. From (\ref{Bound_3FAux_1}) these  boundary conditions translate into corresponding boundary conditions for $Y^A$.}

The Lagrangian \eqref{LY} has been studied at length in reference \cite{Groh:2012tf}, to which we refer for further details. In \cite{Groh:2012tf}  the boundary term has been identified by requiring a consistent variational principle (for previous discussions in non-supersymmetric settings see  
e.g.\ \cite{Brown:1987dd,Duff:1989ah}). On the other hand,  the boundary term is automatically  produced by our duality procedure, once we fix the form of the Lagrangian \eqref{LPhiXr}. The only apparent ambiguity, related to the choice of the form $\ii\int \d^4\theta  (X_A-\bar X_A)U^A$ of the Lagrange multiplier term  in \eqref{LPhiXr},  is completely fixed by the following criterion: (\ref{X+bX}) must be produced without having to impose specific boundary conditions for the gauge superfield $U^A$. Combined with the boundary condition $\delta\Phi^A|_{\rm bd}=0$, this implies that in the dual theory \eqref{LY} we need only impose the gauge invariant boundary condition
\begin{equation}\label{Bound_3F_Condb}
\delta Y^A\big|_\text{bd}= \frac \ii 4 \Big( \bar{D}^2 \delta U^A \Big) \big|_{\mathrm{bd}} = 0\,.
\end{equation}

As a simple consistency check of the equivalence between the Lagrangians (\ref{LY})  
and \eqref{LPhir} we calculate 
the variation of \eqref{LY} with respect to $U^A$,
which results in an equation of motion of the form 
\be
\label{EOMLLBB}
\Im\left(- \frac14 \bar D^2 K_A + \hat W_A\right) = 0 \, . 
\ee
Combining this equation with the (anti)chirality of its components, it follows that
\be
- \frac14 \bar D^2 K_A + \hat W_A=r_A,
\ee
where $r_A$ can be identified with the  real constants appeared in \eqref{LPhir}.

Finally, let us present the explicit form of the bosonic sector of the dual Lagrangian:
\be
\label{LagThreeFormComp}
\begin{split}
\call^{\rm bos}= & K_{A \bar B}\left(D^{A}- \ii \partial_m A^{mA} \right) \left(D^{ B}+ \ii \partial_n A^{n B} \right)
\\
& + \left[ \ii \hat{W}_A \left(D^A-\ii \partial_m A^{Am} \right) + \text{c.c.}\right] + \call^{\rm bos}_{\rm bd}\, ,  
\end{split}
\ee
with
\be
\label{compbound}
\begin{split}
\call^{\rm bos}_{\rm bd} = & - \partial_m \left[ \ii A^{mA}  \left( K_{B \bar A} - K_{A \bar B} \right) D^B 
+ A^{mA}  \left( K_{B \bar A} + K_{A \bar B} \right)  \del_n A^{Bn} 
\right] 
\\
& - \del_m  \left( A^{mA} \hat W_A + A^{mA} \bar{\hat{W}}_{\bar A} \right) \, ,  
\end{split}
\ee
where $A^{Am}\equiv\frac 1{3!}\varepsilon^{mnlp}A^A_{nlp}=({}^*\!A^A_3)^m$. 
Notice that the boundary term automatically guarantees a consistent variational principle.

\subsection{Double three-form multiplets}\label{2-4}

Let us now consider the dualization of a  Lagrangian with a slightly more general superpotential \eqref{sup0} in which $\calg_{AB}$ is a  generic  {\em constant} matrix and its imaginary part is invertible,  $\det(\Im\calg_{AB})\neq 0$.  Hence we can introduce  the arbitrary {\em complex} constants
\be
c_A\equiv e_A+\calg_{AB}m^B \, , 
\ee
and rewrite the Lagrangian in the form
\be\label{LPhi}
{\mathcal L}=\int \d^4\theta\, K(\Phi,\bar\Phi)+\Big(\int \d^2\theta \big[c_A\Phi^A+\hat W(\Phi)\big]+c.c.\Big)\,.
\ee

As in Section \ref{1-4}, we can promote the constants to
chiral superfields $X_A$ by adding appropriate Lagrange multiplier terms to the Lagrangian. The modified Lagrangian is
\be\label{LPhiX}
\call'=\int \d^4\theta\, K(\Phi,\bar\Phi)+\Big(\int \d^2\theta \Big[X_A\Phi^A +\frac14 \bar D^2(\bar X_{\bar A}\Sigma^{\bar A})\Big]+\int\d^2\theta\,\hat W(\Phi)+\text{c.c.}\Big)
\ee
where $\Sigma^{\bar A}$ are complex linear multiplets,   i.e.\
complex scalar superfields satisfying the constraint 
 \be\label{U}
\bar D^2 \Sigma^{\bar A}=0\,,
\ee
see eq.~\eqref{Sigmaexp} for the component expansion of $\Sigma$.
This constraint is explicitly solved in terms of a general Weyl spinor superfield $\Psi^A_\alpha$ as 
\be\label{U=DPsi}
\Sigma^{\bar A}=\bar D\bar\Psi^{\bar A}\,.
\ee
By integrating out $\Psi^A_\alpha$ from (\ref{LPhiX}) we get the condition $D_\alpha X_A=0$ which, combined with the chirality of $X_A$, implies that 
\be\label{X=c}
X_A=c_A,
\ee
where $c_A$ are arbitrary constants. 
Inserting \eqref{X=c} 
into (\ref{LPhiX}) one gets back the Lagrangian (\ref{LPhi}). On the other hand, we can integrate out $X_A$ by imposing their equations of motion and get
\be\label{deltaXbar}
\Phi^A= S^A\equiv - \frac 14 \bar D^2 \bar\Sigma^A\,,
\ee
where $S^A$
are chiral superfields with the following $\theta$-expansion 
\be\label{Si}
 S^A=s^A+\sqrt{2}\theta\lambda^A+\theta^2\, {}^*\!G^A_4 \,.
 \ee
Here $ {}^*\!G^A_4$ are  Hodge duals\footnote{In our conventions, the four-dimensional Hodge dual of a $p$-form $\omega$ is defined by $({}^*\!\omega)_{m_1\ldots m_{4-p}}=\frac{1}{p!}\varepsilon_{m_1\ldots m_{4-p} n_1\ldots n_p}\omega^{n_1\ldots n_p}$, where $\varepsilon^{0123}=-\varepsilon_{0123}=1$.} of the field-strengths
\be
G^A_4=\d C_3^A \, , 
\ee
of complex 3-form gauge fields $C^A_3$. The Hodge duals of $C^A_3$ are complex vector components of the complex linear superfields $\Sigma^A$ (see Appendix \ref{B}).

We call the chiral superfields $S^A$ {\em double} three-form 
multiplets. These kinds of multiplets were introduced in \cite{Gates:1980az} and considered in more detail in \cite{Nishino:2009zz} but, in contrast to the single three-form multiplets $Y^A$ of Section \ref{1-4}, they have attracted much less attention in the literature. 
The bosonic gauge transformation $C_3^A\rightarrow C_3^A+\d\Lambda^A_2$ (where $\Lambda^A_2$ is a complex two-form) are part of the gauge superfield transformation 
\be\label{delta S}
\Sigma^A \quad \to \quad \Sigma^A+L^A_1+\ii L^A_2 \,,
\ee
where $L^A_1$ and $L^A_2$ are  real linear superfields.

It is easy to see that (\ref{delta S}) leaves  $S^A$  invariant. The counterparts of the gauge transformations \eqref{delta S} acting on the `prepotential' $\Psi_\alpha$ are
\be\label{delta Psi}
\Psi^A_\alpha\quad  \to \quad \Psi^A_\alpha+\Lambda^A_\alpha+D^\beta\Lambda^A_{\beta\alpha}\,,
\ee
where $\bar D_{\dot\beta}\Lambda^A_\alpha=0$ and $\Lambda^{A}_{\beta\alpha}=\Lambda^{A}_{\alpha\beta}$.

Note that the Lagrange multiplier term in \eqref{LPhiX} is singled out by a criterion analogous to the one introduced at the end of Section \ref{1-4}. Namely, it leads to (\ref{X=c}) without the need for  any specific boundary condition on the gauge superfield $\Psi_\alpha^A$ and it directly gives back the original Lagrangian, without involving possible boundary terms. As a consequence, the dual Lagrangian describing the dynamics of the superfields $S^A$ is also completely fixed, including the appropriate  boundary term. Indeed, by plugging  \eqref{deltaXbar} back  into \eqref{LPhiX}, we get the dual Lagrangian  
\be\label{LS}
\hat{\mathcal L}=\int \d^4\theta\, K(S,\bar S)+\Big(\int \d^2\theta \,\hat W(S)+\text{c.c.}\Big)+\mathcal L_{\mathrm{bd}}\,,
\ee 
where the boundary term is given by the following total derivative contribution to the Lagrangian
\be\label{bound21}
\mathcal L_{\mathrm{bd}}=\frac14\left(\int\d^2\theta\bar D^2-\int\d^2\bar\theta  D^2\right)\left(\bar X_{\bar A} \Sigma^{\bar A}\right)
+\text{c.c.} \, . 
\ee
In \eqref{bound21} $X_A$ should be replaced by its expression obtained from (\ref{LPhiX}) as the equation of motion of $\Phi^A$, namely
\be\label{deltaPh}
X_A=\frac 14 \bar{D}^2 K_{A}-W_A\,.
\ee
An example of the component field form of the boundary term which one gets from \eqref{bound21} is given in Appendix \ref{A}.\footnote{The free Lagrangian $\hat\call_{\rm free}=\int \d^4\theta  S\bar S$ was briefly discussed in \cite{Gates:1980az}. 
The component form of \eqref{LS} 
with $K=\delta_{A\bar B}S^A\bar S^{\bar B}$ and $\hat W(S)=m_{AB}S^AS^B+g_{ABC}S^AS^BS^C$ but without the boundary term was considered in \cite{Nishino:2009zz}.} 

Let us now turn to the case of constant $\calg_{AB}$ with {\em non}-invertible imaginary part $\Im\calg_{AB}$. If $A,B=1,\ldots,n$, then the matrix $\Im\calg_{AB}$ has a rank $r<n$. This implies that there are $n-r>0$ vectors $u^A_a$, $a=1,\ldots,n-r$, such that $\Im\calg_{AB}\,u^B_a=0$. We can complete them with $r$ vectors $v^A_q$, $q=1,\ldots,r$, which together with $u^A_a$ form a basis of $\mathbb{R}^{n}$. We can use this basis to re-organize the chiral superfields as follows 
\be\label{Phisplit}
\Phi^A=\Phi^a u^A_a+\Phi^q v^A_q\,.
\ee
and, analogously, $m^A= m^a u_a^A+ m^p v_p^A$.
Then the superpotential (\ref{sup0}) takes the form
\be
W=r_a\Phi^a+c_p\Phi^p+\tilde W(\Phi),
\ee
where
\be\label{rcdef}
r_a\equiv (e_A +m^B\Re\calg_{AB})u^A_a\,,\quad c_p\equiv e_Av^A_p + m^qv^A_q\calg_{AB}v^B_p
\ee
are, respectively, arbitrary real and complex constants and
\be\label{W'}
\tilde W\equiv \hat W+m^a u_a^A\calg_{AB}v^B_p\Phi^p.
\ee

We can then proceed by dualizing $\Phi^a$ to single three-form multiplets $Y^a$ as in Section \ref{1-4} and $\Phi^q$ to double three-form multiplets $S^q$ as in the present Section.\footnote{Notice that the choice of the vectors $u^A_a$ is not unique, as we could redefine $v^A_q\rightarrow v^A_q+\alpha_q^a u^A_a$ with $\alpha_q^a$ being arbitrary real constants. This ambiguity induces the redefinitions  $\Phi^a\rightarrow \Phi^a-\alpha_q^a\Phi^q$ and $c_q\rightarrow c_q+\alpha_q^a r_a$, which  mix the two kinds of dual three-form multiplets.}

\subsection{Double three-form multiplets and non-linear dualization}\label{nld}

We are now ready to consider the more general case of non-constant holomorphic matrix $\calg_{AB}(\Phi)$, still in the case of rigid supersymmetry. Even though not explicitly indicated, the following discussion allows for the inclusion of additional chiral multiplets  in the theory, which can  enter $\calg_{AB}(\Phi)$ and $\hat W(\Phi)$ in (\ref{sup0}), but which are not subject to the dualization procedure. For instance, extra  chiral multiplets $T^p$ will explicitly appear in Section \ref{sec:IIA},  in which we will apply our construction to  type IIA flux compactifications.

For convenience we define the matrices
\be\label{splitgAB}
\caln_{AB}=\Re\calg_{AB}\,,\quad\calm_{AB}=\Im\calg_{AB}.
\ee
We will assume that, for generic values of the chiral fields $\Phi^A$, the matrix $\calm_{AB}$ is invertible.  We will briefly come back to the degenerate case $\det(\calm_{AB})=0$ at the end of the section. Furthermore, for simplicity, we assume  that $\calg_{AB}(\Phi)$ is symmetric, although most of the discussion holds for non-symmetric $\calg_{AB}(\Phi)$. This symmetry  is automatic if we regard $\calg_{AB}(\Phi)$ as the second derivative of a holomorphic prepotential $\calg(\Phi)$, as we will assume in the local supersymmetry case.

Our starting point is the Lagrangian 
\be\label{callgen}
\call=\int\d^4\theta\, K(\Phi,\bar\Phi)+\left(\int\d^2\theta\left[e_A\Phi^A+m^A\calg_{AB}(\Phi)\Phi^B+\hat W(\Phi)\right]  
+\text{c.c.}\right) \, . 
\ee
The strategy followed in the previous Sections  is then generalized by replacing \eqref{callgen} with the following Lagrangian
\be\label{callgen'}
\begin{aligned}
\call'' =&\int\d^4\theta\, K(\Phi,\bar\Phi)+\left(\int\d^2\theta\left[X_A\Phi^A+\hat W(\Phi)\right]+\text{c.c.}  \right)\\
&-\frac14\left(\int\d^2\theta\, \bar D^2\left[\Sigma_A\,\calm^{AB}(X_B-\bar X_B)\right]+\text{c.c.}\right)
\end{aligned}
\ee
where $\calm^{AB}$ is the inverse of $\calm_{AB}$ and $\Sigma_A=\bar D\bar\Psi_A$ are complex linear superfields defined by eqs. \eqref{U} and \eqref{U=DPsi}.

The extremization  of (\ref{callgen'}) with respect to $\Psi^\alpha_A$ gives 
\be\label{XAeq}
D_\alpha(\calm^{AB}\,\Im X_B)=0\,. 
\ee
Notice that the variation of \eqref{callgen'} with respect to $\Psi^A$ does not involve any boundary terms and the Lagrange multiplier term in \eqref{callgen'} satisfies the criterion discussed in the previous Sections. The general solution of \eqref{XAeq} is 
\be\label{vevX_A}
X_A =e_A+\calg_{AB}(\Phi)m^B \, , 
\ee
with  $e_A$ and $m^B$ being arbitrary real constants.\footnote{Indeed, from (\ref{XAeq}) and its complex conjugate one gets $\calm^{AB}\Im X_B=m^A$, with $m^A$ being arbitrary real constants.  We can then write $ X_{A}=\Re X_A+\ii{\rm Im}\, \calg_{AB}\,m^B\equiv \Re (X_A-\calg_{AB}m^B)+\calg_{AB}m^B$. 
This equation is compatible with the chirality of $X_A$ and $\calg_{AB}$ only if $\Re (X_A-\calg_{AB}m^B)=e_A$ are constant. We thus arrive at (\ref{vevX_A}).} Hence, by plugging (\ref{vevX_A}) back into (\ref{callgen'}) one obtains the original Lagrangian  (\ref{callgen}).

Alternatively, we  get the dual description by integrating out $X_A$ in (\ref{callgen}). 
This results in the following expression for the chiral superfields $\Phi^A$ 
\be\label{genPhi}
\Phi^A = S^A\,,
\ee
where
\be\label{nLS}
S^A\equiv\frac 14 \bar D^2\left[\calm^{AB}(\Sigma_B-\bar\Sigma_B)\right]\,.
\ee
The chiral superfields $S^A$ provide a generalization of the double three-form multiplets encountered in Section \ref{2-4}. Note that, once we impose (\ref{genPhi}),  $\calm^{AB}$ depends  on $S^A$. Then, in general, the expression (\ref{nLS})  is non-linear and  cannot be explicitly solved for $S^A$ as a function of  $\Sigma_A$. 
However, this does not necessarily create complications in specific applications, 
as for instance to type IIA flux compactifications discussed in Section \ref{sec:IIA}.

The above formulation in terms of $\Sigma_A$, that contains gauge three-forms, is invariant under the following gauge transformations which generalize \eqref{delta S}
\be
\label{GINV}
\Sigma_A\quad\rightarrow \quad \Sigma_A+\tilde L_A+\calg_{AB}L^B\,,
\ee
where $\tilde L_A$ and $ L^B$ are arbitrary real linear superfield parameters. This gauge symmetry guarantees that the gauge three-forms enter \eqref{genPhi} via their gauge-invariant field strengths only. We will discuss the component structure of the relation \eqref{nLS} in the supergravity case in Section \ref{sec:gensugra}. 

If we substitute the solution \eqref{genPhi} back into the Lagrangian \eqref{callgen'} we obtain 
\be\label{LSX}
\hat{\mathcal L}=\int \d^4\theta\, K(S,\bar S)+\Big(\int \d^2\theta \,\hat W(S)+\text{c.c.}\Big)+\mathcal L_{\mathrm{bd}}\,,
\ee 
where the boundary term is now given by the total derivative contribution 
\be\label{bound2}
\mathcal L_{\mathrm{bd}}=\int\d^2\theta\left(\int\d^2\bar\theta+\frac14\bar D^2\right)\left(X_A\calm^{AB}(\Sigma_B-\bar\Sigma_B)\right)
+\text{c.c.} \, , 
\ee 
in which $X_A$ is expressed via $\Sigma_A$ on account of the equation of motion of $\Phi^A$, as in Section \ref{2-4}.
We will give the explicit expression of the boundary term in the supergravity case in the next Section.  

The Lagrangian  (\ref{LS}) provides us with the dual formulation of the considered theory in terms of the double three-form multiplet \eqref{nLS}, with the `reduced' superpotential $\hat W(S)$ and the appropriate boundary term. The information about the form of the matrix $\calg_{AB}(\Phi)$ appearing in the superpotential of the original theory is encoded in the form of the matrix $\mathcal M^{AB}$ which enters the definition (\ref{nLS}) of the double three-form multiplet. On the other hand, as in the previous Sections, the constant parameters $e_A$ and $m^A$ got dualized into the expectation values of the field-strengths of the gauge three-forms.

Before passing to the locally supersymmetric case, let us briefly discuss the situation in which ${\rm Im}\,\calg_{AB}$, with $A,B=1,\ldots,n$, is degenerate of rank $r<n$. Then there should exist $n-r>0$  real vectors $u^A_a$, $a=1,\ldots, n-r$, such that ${\rm Im}\,\calg_{AB}(\Phi)u^B_a=0$ and hence $\calg_{AB}(\Phi)u^B_a=\bar\calg_{AB}(\bar\Phi)u^B_a$. Taking into account the holomorphicity of  $\calg_{AB}(\Phi)$, this condition is quite strong and puts strong constraints on the form of $\calg_{AB}(\Phi)$. Suppose, for instance, that the vectors $u^A_a$ are constant, as at the end of Subection \ref{2-4}. This would imply that $\calg_{AB}(\Phi)u^A_a$ is constant too. We could then proceed as in Section \ref{2-4}, rewriting the superpotential as  follows
\be\label{mixedW}
W=r_a\Phi^a+ [e_p+m^q \calg_{qp}(\Phi)]\Phi^p+\hat W'(\Phi),
\ee
where $e_a\equiv e_A u^A_a$, $\calg_{qp}\equiv u^A_q\calg_{AB}u^B_p$, $r_a$ is as in (\ref{rcdef}) and $\hat W'$ is as in (\ref{W'}). One can then
dualize $\Phi^a$ to single three-form multiplets $Y^a$  and $\Phi^p$ to  double three-form multiplets $S^p$. 
We expect similar combinations of different dualizations to be possible in more general cases.

%%%%%%%%%%%%%%%%%%%%%%%%%%%%%%%%%%%%%%%%%%%%%%%%%%%%%%%%%%%%%%%%%%%%%%%%%%%%%%%%%%%%%%%%%%%%%%%%%%%%%%%%%%%%%%%%%%%%%%%%%%%%%%%%%%%%%%%%%%%%%%%%%%
\section{Three-form multiplets in $\mathcal N=1$ supergravity}\label{sugracase}

We now extend to matter-coupled $\caln=1$ supergravity the duality procedure described in Section \ref{rigid} for rigid supersymmetry.
The extension is rather natural if we use a super-Weyl invariant approach \cite{Howe:1978km}. 
Before proceeding let us recall that the old-minimal formulation of supergravity \cite{Stelle:1978ye}
describes the interactions of the gravitational multiplet 
\be
e_a^m \ , \quad \psi_m^\alpha  \ , \quad  b_a  \ , \quad  M \ .  
\ee
The physical fields are the vielbein $e_a^m$ and the gravitino $\psi_m^\alpha$, 
whereas the auxiliary fields are the real vector $b_a$ 
and the complex scalar $M$.

We will construct three-form matter-coupled supergravity by dualizing a super-Weyl invariant formulation. The curved superspace supervielbeins  transform as follows under the super-Weyl transformations \cite{Howe:1978km}
\be\label{EW}
E^a_M\rightarrow e^{\Upsilon+\bar\Upsilon}E^a_M\,,\quad E^\alpha_M\rightarrow e^{2\bar\Upsilon-\Upsilon}
\left(E^\alpha_M-\frac{\ii}{4}E^a_M\sigma^{\alpha\dot\alpha}_a\bar\cald_{\dot\alpha}\bar\Upsilon\right).
\ee
where $(a,\alpha)$ are flat superspace indices, $M=(m,\mu)$ are curved indices and $\Upsilon$ is an arbitrary chiral superfield parameterizing the super-Weyl transformation.
We will focus on a theory for $n+1$ chiral multiplets  $\calz^A$, $A=0,\ldots,n$,  that transform as follows under super-Weyl  transformations
\be\label{weylZ}
\calz^A\rightarrow e^{-6\Upsilon}\calz^A.
\ee
The chiral superfields $\calz^A$ comprise,  in a democratic way, a super-Weyl compensator and $n$ physical multiplets. 

The ordinary old-minimal formulation of supergravity is obtained by choosing a super-Weyl compensator $Z$, e.g.\ $Z\equiv {\cal Z}^0$, and subject it to a gauge-fixing condition using the super-Weyl invariance. On the other hand, we will perform the duality transformation of the conventional chiral multiplets $\calz^A$  to three-form multiplets {\em before} gauge-fixing the  super-Weyl invariance.  In this way, the procedure will work exactly as in the rigid supersymmetry case, but will involve the super-Weyl compensator in addition to the physical chiral superfields. Gauge-fixing the super-Weyl symmetry afterwards will produce a Lagrangian describing the coupling of three-form multiplets to a supergravity multiplet with one or two gauge fields substituting the scalar auxiliary fields.

In the next Section we will focus on pure supergravity and its three-form variants. 
The inclusion  of additional physical chiral multiplets and a general superpotential of the form (\ref{sup0}) will be considered in Section \ref{sec:gensugra}. The following discussion can include additional `spectator' matter or gauge multiplets, which will not be explicitly indicated for notational simplicity.

%%%%%%%%%%%%%%%%%%%%%%%%%%%%%%%%%%%%%%%%%%%%%%%%%%%%%%%%%%%%%%%%%%%

\subsection{Variant minimal supergravities from  duality}
\label{sec:varmin}

We start by considering the  minimal theory, in which the old-minimal supergravity multiplet is coupled just to the super-Weyl compensator $Z$, which transforms as in (\ref{weylZ}). Then, up to a complex constant rescaling of $Z$, the most general super-Weyl invariant Lagrangian has the form
\be\label{SWL}
\mathcal L=-3\int \d^4\theta \, E\, (Z\bar Z)^{\frac 13} + \left( c\int \d^2\Theta \, 2 {\cal E}\,Z  + \text{c.c.} \right) \, , 
\ee
in which $E$ denotes the Berezinian super-determinant of the super-vielbein, $ \d^2\Theta \, 2 {\cal E}$ is a chiral superspace measure \cite{Wess:1992cp} and $c$ is an arbitrary complex number which gives rise to the gravitational cosmological constant and the gravitino mass. Under  (\ref{EW}), the superspace measures re-scale as 
\be
E \to e^{2(\Upsilon+\bar\Upsilon)}\,E \ , \qquad \d^2\Theta\,\mathcal E\to  e^{6\Upsilon}\,\d^2\Theta\,\mathcal E \, .
\ee
Hence the  super-Weyl invariance of the supergravity Lagrangian is manifest. 
We can now follow Section \ref{rigid}, distinguishing two cases.

\subsubsection{ Single three-form supergravity}
\label{sec:min1}

We first  proceed along the lines of Section \ref{1-4}, setting $c\equiv \ii r$, with real $r$, and promoting $r$ to a chiral multiplet $X$ by adding  an appropriate Lagrange multiplier.\footnote{We choose a purely imaginary $c$ in order to obtain the single three-form supergravity in its most common form, as used for instance in \cite{Gates:1980az,Buchbinder:1988tj,Ovrut:1997ur,Farakos:2016hly,Kuzenko:2005wh,Bandos:2011fw,Bandos:2012gz}. Clearly, by a simple redefinition $Z\rightarrow -\ii Z$ one can make $c$ purely real.} 
Consider the modified Lagrangian  
\be\label{SWL1}
\mathcal L'=-3\int \d^4\theta \, E\, (Z\bar Z)^{\frac 13} 
+ \left( \int \d^2\Theta \, 2 {\cal E} \, \left[ X Z + \frac{1}{8}\left( \bar{\cal D}^2 - 8 {\cal R} \right) 
\left\{ U (X + \bar X) \right\} \right] + \text{c.c.} \right) \,  , 
\ee
where  $U$ is a scalar real superfield and  $\cal R$ is the chiral superfield curvature whose leading component is 
the auxiliary 
field $M = - \frac16 {\cal R}|$ of the gravity multiplet. 
Notice that (\ref{SWL1}) is super-Weyl invariant if we impose that 
\be\label{Uweyl}
U\to e^{-2(\Upsilon+\bar\Upsilon)}U \, , 
\ee 
under super-Weyl transformations, since  $\bar{{\cal D}}^2-8{\mathcal R}\to e^{-4\Upsilon}(\bar{{\cal D}}^2-8{\mathcal R})  e^{2 \bar{\Upsilon} }$.

Integrating $U$ out of (\ref{SWL1}) by imposing  its equation of motion implies that $X$ must be  
an arbitrary real constant $r$ and then one goes back to (\ref{SWL}). Instead, integrating out  $X$ gives
\be\label{rZ}
Z\equiv Y\,,
\ee
where the chiral superfield
\be\label{YY}
Y\equiv -\frac14 \left( \bar{\cal D}^2 - 8 {\cal R} \right)U\, 
\ee
is the natural generalization of the rigid single three-form multiplets discussed in Section \ref{1-4}. 
In particular, the bosonic three-form $A_3$ is contained in the component 
\be\label{A=DDU} 
- \frac18 \bar \sigma_m^{\dot \alpha \alpha} [{\cal D}_\alpha , \bar{\cal D}_{\dot \alpha} ] U \Big{|}   
\equiv ({}^*\!A_3)_m\, , 
\ee
of $U$. The bosonic gauge transformation $A_3\to A_3+\d\Lambda_2$ is contained in the superfield gauge transformation $U\rightarrow U+L$, 
where $L$ is an arbitrary linear multiplet.  
This gauge invariance allows one to write the superfield $U$ in an appropriate WZ gauge $U|=0$, 
which we have already used in \eqref{A=DDU}.

By integrating out $Z$ one gets the equation
\be\label{Xeq}
X=  - \frac14 \left( \bar{\cal D}^2 - 8 {\cal R} \right) \left[ Z^{-\frac23} \bar Z^{\frac13}  \right] \, , 
\ee
and by plugging (\ref{rZ}) and (\ref{Xeq}) back into (\ref{SWL1}) one obtains  the dual Lagrangian
\be\label{hatLYY}
\hat \call=-3\int \d^4\theta \, E\, (Y\bar Y)^{\frac 13} +\call_{\rm bd},
\ee
where
\be
\call_{\rm bd}= \frac18 \int \d^2\Theta \, 2 {\cal E} \, \left( \bar{\cal D}^2 - 8 {\cal R} \right) 
\left\{ U (X - \bar X) \right\}  + \text{c.c.}  
\ee 
Note that $\call_{\rm bd}$ is indeed a total derivative. 
$Y$ transforms  as $Z$ under super-Weyl transformations
$
(Y\rightarrow e^{-6\Upsilon}Y)
$
and plays the role of the super-Weyl compensator. 

It is known that different off-shell formulations of four-dimensional $\mathcal N=1$ supergravity can be obtained from its superconformal  version by choosing different  compensator fields \cite{Gates:1983nr,Buchbinder:1995uq,Freedman:2012zz}. 
Here the use of $Y$ as a compensator in the super-Weyl invariant formulation 
leads, as was shown in \cite{Kuzenko:2005wh}, to the three-form minimal supergravity  \cite{Gates:1980az,Buchbinder:1988tj,Ovrut:1997ur,Bandos:2011fw,Bandos:2012gz,Farakos:2016hly}, 
in which the imaginary part of the old-minimal auxiliary field $M$ is substituted 
by the Hodge dual of a real field-strength $F_4=\d A_3$.

In order to see this, we can  use the super-Weyl symmetry  to set 
\be\label{rZ1}
Y=1\,.
\ee 
By recalling the definition of $Y$ given in  \eqref{YY}, its expansion (\ref{Y}) and skipping the dependence on the fermions, the lowest component of this equation gives
$Y|=1$ while the highest component $-\frac 14 \mathcal D^2 Y|=0$ gives $\Im M+{}^*\!\d A_3=0$, so that
the conventional scalar auxiliary field of the supergravity multiplet has the form $M = \Re M -\ii {}^*\!F_4$, as proposed in \cite{Gates:1980az} and discussed in detail in \cite{Ovrut:1997ur}. 
Hence, the component fields of the supergravity multiplet of this formulation are 
\be
e_a^m \ , \quad \psi_m^\alpha  \ , \quad b_a  \ , \quad M_0  \ , \quad A_3 \  , 
\ee 
where $M_0\equiv \Re M$ is a real scalar.

\subsubsection{Double three-form supergravity}
\label{sec:min2}

In order to arrive at the minimal double three-form supergravity \cite{Stelle:1978ye} we must promote the entire arbitrary constant $c$ to a dynamical chiral field $X$ and proceed as in the previous examples. This can be done by starting from the Lagrangian
\be\label{SWL2}
\mathcal L'=-3\int \d^4\theta \, E\, (Z\bar Z)^{\frac 13} 
+ \left( \int \d^2\Theta \, 2 {\cal E}\, 
\left[ 
X Z +\frac{1}{4} \left( \bar{\cal D}^2 - 8 {\cal R} \right) 
\left\{ \bar X \Sigma \right\}  
\right] 
+  \text{c.c.} \right) \, , 
\ee
where $\Sigma=\bar\cald\bar\Psi$ is a complex linear superfield,  the locally supersymmetric generalization of  the complex linear superfield introduced in Section \ref{2-4}. 
The components of $\Sigma$ in the appropriate WZ gauge are 
\begin{equation}
\begin{aligned}
\Sigma | &  = 0 \, , 
\\
{\cal D}^2 \Sigma | &  = -4 \bar s \, ,
\\
\frac12\bar\sigma_m^{\dot \alpha \alpha} [{\cal D}_\alpha , \bar{\cal D}_{\dot \alpha} ] \Sigma |& 
= - \ii C_m \, , 
\\
{\cal D}^2 \bar{\cal D}^2 \bar \Sigma |& = 
8 
{\,}^* \!\bar G_4 
+ 16 \bar M s \, ,
\end{aligned}
\end{equation}
with $G_4\equiv \d C_3$ and $C_m\equiv ({}^*\!C_3)_m$. One can go to this gauge because of the invariance of the construction under the 
superfield gauge transformation of the form (\ref{delta S}) - (\ref{delta Psi}).

The action (\ref{SWL2}) is invariant under super-Weyl transformations if $\Psi_\alpha$, 
and eventually $\Sigma$, 
transform as follows \cite{Buchbinder:1995uq} 
\be
\Psi_\alpha \to e^{-3 \Upsilon}\Psi_\alpha \  , \qquad \Sigma\to e^{-2(\Upsilon+\bar\Upsilon)}\Sigma \, . 
\ee
As in the previous examples, by integrating out $\Psi_\alpha$ one gets back (\ref{SWL}). 
On the other hand, by integrating out $X$ and $Z$ one finds 
\be
\label{ZandX}
\begin{split}
Z & =  S \equiv-\frac 14\left( \bar{\cal D}^2 - 8 {\cal R} \right)\bar\Sigma \, , 
\\
X & = - \frac14 \left( \bar{\cal D}^2 - 8 {\cal R} \right) \left[ Z^{-\frac23} \bar Z^{\frac13}  \right] \, . 
\end{split}
\ee 
After inserting these expressions into the Lagrangian 
one arrives at 
the dual description 
\be
\label{anotherone}
\hat{\mathcal L}=-3\int \d^4\theta \, E\, (S\bar S)^{\frac 13} 
+ 
\frac14 \left[
\int \d^2\Theta \, 2 {\cal E} \, \left( \bar{\cal D}^2 - 8 {\cal R} \right) 
\left\{ \bar X \Sigma - X \bar \Sigma \right\}  + \text{c.c.}  \right] 
\, , 
\ee
where $S$ 
is a double three-form multiplet  which plays the role of the super-Weyl compensator. 
Note that $X$ and $S$ in \eqref{anotherone} are given by \eqref{ZandX}, 
and that the second term in \eqref{anotherone} is the boundary term.

One can then gauge-fix the super-Weyl invariance by putting $S=1$ and find that 
\be\label{M=flux}
M = - \frac 12 {}^*\!G_4   
\ee 
Hence the supergravity multiplet in this formulation becomes 
\be
e_a^m \ , \quad \psi_m^\alpha  \ , \quad  b_m  \ , \quad C_3 \ , 
\ee
where $C_3$ is a complex three-form. 
Therefore we refer to this formulation as {\it double} three-form supergravity. 
The bosonic sector of this minimal supergravity  theory follows from the Lagrangian \eqref{anotherone} and has the following form 
\be
\label{another boson sector}
e^{-1} \hat{\mathcal L} = 
- \frac12 R 
+ \frac13 b^m b_m 
- \frac{1}{12} \big{|} {}^*\!G_4  \big{|}^2 
+  \frac1{12} \cald_m \left( 
C^m \, {}^*\! \bar G_4 
+ \text{c.c.} 
\right) \, .  
\ee 
The equations of motion of $C_3$ have  general solution 
\be\label{G4c}
{}^*\!G_4 = 6\,c  \, . 
\ee
If we integrate out $C_3$ by inserting \eqref{G4c} into the Lagrangian \eqref{another boson sector} we find the standard 
supergravity theory with a negative cosmological constant. 
Notice that \eqref{another boson sector} has a well defined variation with respect to $C_3$ thanks to the presence of the boundary term. As in the previous Sections, this is guaranteed by our duality procedure once one appropriately chooses the form of the Lagrange multiplier term in \eqref{SWL2}.

%%%%%%%%%%%%%%%%%%%%%%%%%%%%%%%%%%%%%%%%%%%%%%%%%%%%%%%%%%%%%%%%%%%

\subsection{Three-form matter-coupled supergravities}
\label{sec:gensugra}

In the previous Section we obtained known minimal three-form supergravities with the use of the locally supersymmetric counterpart of the duality procedure described in Section \ref{rigid}. We now pass to the considerably more general  case outlined at the beginning of this section. 
We consider a super-Weyl invariant supergravity theory coupled to $n+1$ chiral superfields $\calz^A$ which transform as in (\ref{weylZ}). We stress once again that, even if not explicitly indicated for notational simplicity, additional spectator chiral  and vector multiplets may be included without difficulties (as in the example discussed in Section \ref{sec:IIA}).

The general form of the super-Weyl invariant Lagrangian is 
\be\label{NWL}
\call=-3\int \d^4\theta\, E\,\Omega({\mathcal Z},\bar {\mathcal Z}) 
+ \left(\int\d^2\Theta\,2\cale\,\calw(\calz) +c.c \right)\, , 
\ee 
where the kinetic potential $\Omega({\mathcal Z},\bar {\mathcal Z})$ and the superpotential $\calw(\calz)$ have the following homogeneity properties
\be\label{OWhom}
\Omega(\lambda \mathcal Z,\bar\lambda\bar {\mathcal Z})=|\lambda|^{\frac 23}\Omega({\mathcal Z},\bar {\mathcal Z})\,,\qquad 
\calw(\lambda \mathcal Z)=\lambda\calw(\mathcal Z).
\ee
Before discussing the duality procedure, let us briefly recall how this formulation is related to the more standard supergravity formulation. First, one singles out a super-Weyl compensator $Z$ as follows 
\be\label{Zsplit}
 \calz^{ A}=Z \calz_0^A(\Phi) \, , 
\ee  
where ${\mathcal Z}_0^{A}(\Phi)$ is a set of functions of the physical chiral multiplets $\Phi^i$  $(i=1,\ldots,n)$, which are inert under the super-Weyl transformations. Clearly, the split (\ref{Zsplit}) has a large arbitrariness and one may redefine
\be\label{arb}
Z\rightarrow e^{-f(\Phi)}Z\,,\quad \mathcal Z_0^A(\Phi)\rightarrow e^{f(\Phi)} \mathcal Z_0^A(\Phi)\,.
\ee
The kinetic potential $\Omega({\mathcal Z},\bar {\mathcal Z})$  can be written as follows
\be
\Omega({\mathcal Z},\bar {\mathcal Z})=|Z|^{\frac23}e^{-\frac13 K(\Phi,\bar\Phi)},
\ee
where $K(\Phi,\bar\Phi)\equiv -3\log \Omega({\mathcal Z}_0(\Phi),\bar {\mathcal Z}_0(\bar\Phi))$ is the ordinary K\"ahler potential.
Note that the possibility of making the redefinition (\ref{arb}) corresponds to the  invariance under K\"ahler transformations 
$ K(\Phi,\bar\Phi)\rightarrow K(\Phi,\bar\Phi)-f(\Phi)-\bar f(\bar\Phi)$. The conventional superpotential $W(\Phi)$ is singled out by using the split (\ref{Zsplit}) and defining
\be\label{WWrel}
\calw(\mathcal Z)= Z\,W(\Phi)\,,
\ee
where $W(\Phi)\equiv \calw(\mathcal Z_0(\Phi))$. Under the redefinition (\ref{arb}) $W$ transforms as follows $W(\Phi)\rightarrow e^{f(\Phi)}W(\Phi)$. The conventional formulation can then be obtained by gauge-fixing the super-Weyl invariance, e.g. by putting
\be\label{Zgfix}
Z=1\,.
\ee

In order to perform the duality procedure, let us come back to the super-Weyl invariant Lagrangian (\ref{NWL}) and consider the superpotential of the form
\be\label{restrW}
\calw(\calz)\equiv e_A\calz^A+m^B \calg_{BA}(\calz)\calz^A +\hat\calw(\calz).
\ee 
The homogeneity condition (\ref{OWhom}) requires that $\calg_{AB}(\lambda\calz)=\calg_{AB}(\calz)$ and $\hat\calw(\lambda\calz)=\lambda\hat\calw(\calz)$. Though the construction under consideration can be applied to generic  $\calg_{AB}$, we will restrict ourselves to the case in which 
\be
\calg_{AB}(\calz)\equiv \partial_A\partial_B \calg(\mathcal Z) \, , 
\ee
with $\calg({\mathcal Z})$ being  a (possibly locally defined) homogeneous prepotential of degree two $\calg(\lambda {\mathcal Z})=\lambda^2\calg( {\mathcal Z})$ defining a local special K\"ahler space parametrized by homogeneous coordinates
${\mathcal Z}^{A}$, ${ A}=0,1,\ldots,n$.\footnote{The minimal supergravities considered in Section \ref{sec:varmin} correspond to the simplest subcases with $n=0$, $\calz^0=Z$ and $c\equiv e_0+\calg_{00}m^0$, where $\calg_{00}$ is necessarily constant by homogeneity. In particular, the single three-form minimal supergravity of Section  \ref{sec:min1} is obtained by setting $\calg_{00}=0$ and redefining $Z\rightarrow \ii Z$.} As we will see, string flux compactifications have superpotentials of this kind with $(e_{ A},m^{B})$ representing appropriately quantized units of fluxes.

We would like to make the $2n+2$ constants $(e_{A},m^{A})$ in \eqref{restrW} dynamical, i.e. to replace them with the field-strengths of $2n+2$ three-forms.  This is achieved by  dualizing the chiral fields $\mathcal Z^{ A}$, easily adapting the procedure introduced in Section \ref{nld} for the rigid supersymmetric case. As in that Section, we assume that $\calm_{AB}$ defined as in (\ref{splitgAB}) is invertible. (The case  of degenerate $\calm_{AB}$ can be addressed as outlined in Section \ref{rigid}, combining dualizations to single- and double-three form multiplets.)  First, we substitute the chiral superspace integral of the superpotential term (\ref{restrW}) with
\be
\label{NLboundary}
\mathcal L_X=2\int \d^2\Theta \,\cale\,\left (X_{A}\mathcal Z^{ A} - \frac14 \left(\bar{\cal D}^2 - 8 {\cal R}\right) \left [
{\cal M}^{AB} \left(  X_A - \bar X_A \right) \Sigma_B \right]+\hat\calw(\calz)\right)\,,
\ee
where, as in the rigid supersymmetry case,  $\mathcal M^{AB}$ is inverse of $\calm_{AB}={\rm Im}\,\calg_{AB}$, $X_{A}$ 
are chiral superfields and $\Sigma_A$ are complex linear superfields  
$\Sigma_A\equiv\bar{\cal D}_{\dot \alpha} \bar\Psi_{A}^{\dot \alpha}$.  
Upon integrating out $\Psi^\alpha_{ A}$ 
one gets back (\ref{NWL}). On the other hand, by integrating out  $X_{ A}$ and $\calz^A$ one 
finds 
\be
\calz^A = S^{ A} \, , 
\ee
where 
the chiral superfields $ S^{ A}$ are double three-form multiplets, defined by the generalization of \eqref{nLS},
\be\label{mZA} 
 S^{ A}= \frac14(\bar\cald^2-8\calr)\left[\calm^{ {AB}}(\Sigma_{ B}-\bar\Sigma_{ B})\right] \, , 
\ee 
and 
\be
\label{XXAA}
X_A = 
- \hat\calw_A 
+ \frac14 (\bar {\cal D}^2 - 8 {\cal R}) 
\left[ \Omega_A + \frac{\partial \calm^{BC}}{\partial S^A} 
\left(X_B - \bar{X}_B \right) \left( \Sigma_C - {\bar\Sigma}_C \right) \right] \, . 
\ee 
The Lagrangian then reads 
\be\label{OL1}
\hat{\mathcal L}=-3\int\d^4\theta\,E\,\Omega(S,\bar {S})
+\left(\int\d^2\Theta\,2\cale\, \hat\calw( S)+\text{c.c.}\right)+{\mathcal L}_{\rm bd} \, , 
\ee
in which the boundary term is given by the $X$-dependent part of \eqref{NLboundary} once one replaces $\calz^A$ with $S^A$ and $X_A$  with \eqref{XXAA}. 
Note that, as in the rigid supersymmetry case, the dual Lagrangian does not have the part of the superpotential that depended on $e_A$ and $m^A$. 
We thus end up with a theory in which the only independent superfields are the complex linear multiplets $\Sigma_A$.

The double three-form multiplets $S^A$ are defined by (\ref{mZA}), in which ${\cal M}^{AB}$ should be considered as a function of $S^A$ and $\bar S^A$. Hence (\ref{mZA})  is non-linear and  so is not generically solvable for $S^A$ as functions of $\Sigma_A$. However, it  turns out to be tractable for superfield components. For simplicity, we will restrict ourselves to the bosonic ones setting the fermionic components equal to zero.    
Using the local symmetry \eqref{GINV} we can impose the Wess--Zumino gauge in which, in particular, $\Sigma_A|=0$. Then the remaining bosonic components of $\Sigma_A$ are 
\begin{equation}\label{Sigmacomphelper}
\begin{aligned}
{\cal D}^2 \Sigma_A | &  = -4 \bar s_A \, ,
\\
\frac12 \bar\sigma_m^{\dot \alpha \alpha} [{\cal D}_\alpha , \bar{\cal D}_{\dot \alpha} ] \Sigma_A |& 
= - \tilde A_{Am} - 
{\cal G}_{AB} \, A^{B}_m  \, , 
\\
{\cal D}^2 \bar{\cal D}^2 \bar \Sigma_{ A} |& = 
8 \ii \,  {\cal D}_m \left( \tilde A^m_A 
+  \bar{\cal G}_{AB} \, A^{B m} \right)  
+ 16 \bar M s_{A} \, , 
\end{aligned}
\end{equation}
where $A^A_{m}\equiv ({}^*\! A^A_{3})_m$ and $\tilde A_{Am}\equiv ({}^*\! \tilde A_{3A})_m$. 

From \eqref{mZA} it follows that the scalar component $ s_A$, with lower indices, appearing in \eqref{Sigmacomphelper} is  related  to  $ s^A\equiv S^A|$, with upper indices, by the inverse metric $\calm^{AB}$. Since  $S^A|\equiv \calz^A|\equiv z^{A}$, we can use $z^A$ instead of $s^A$ and write this relation as follows 
\be\label{z}
 z^{A} = {\cal M}^{AB} (z,\bar z)\,  s_{B} \,.
\ee
In general it is not possible to explicitly invert the above expression and express $z^A$ in terms of the scalar fields $ s_A$ of the complex linear superfield $\bar \Sigma_A$. Hence, in what follows it will be more convenient to use $z^A$ as independent scalar fields in the component Lagrangians which we will shortly present. The $\theta^2$-component of \eqref{mZA} is then
\be
\label{FA2}
\begin{split}
F^A_S\equiv-\frac 14 {\mathcal D}^2 S^A|= &  \bar M z^{A}  
+ \frac{\ii}{2} {\cal M}^{AB} \left(\,{}^*\!\tilde F_{4B}
+  \bar{\cal G}_{BC} \,{}^*\!F^C_{4} +2\Re\left[\bar{\cal G}_{BCD} \bar{F}_S^D \bar z^C\right] \right)
\, ,
\end{split}
\ee 
where $\tilde F_{4A}=\d\tilde A_{3A}$, $F^A_{4}=\d A^A_{3}$ and ${\cal G}_{A BC}\equiv \partial _A{\cal G}_{ BC}$. 
Now, taking into account that $z^{ A} {\cal G}_{A BC} = 0$ by homogeneity,
we reduce equation \eqref{FA2}  to
\be\label{FgF}
F^{A}_S = \bar M z^{A}  
+ \frac{\ii}{2} {\cal M}^{AB} \left(\,{}^*\!\tilde F_{4B}
+ \bar{\cal G}_{BC} \,{}^*\!F^C_{4}\right) \, . 
\ee
To fix the super-Weyl invariance it turns out to be convenient to choose one of the superfields $S^A$ ($A=(0,i)$), 
say $S^0$, as the super-Weyl compensator and impose 
\be\label{Z=1}
 S^0=1 \, . 
\ee 
The superspace condition \eqref{Z=1} implies the component field conditions $z^0=1$ and $F^0_{S}=0$. 
The bosonic relations \eqref{FgF} split as follows
\be\label{M+F}
\begin{aligned}
\bar M &=-  
\frac{\ii}{2} {\cal M}^{0B}(z,\bar z)\left[ \,{}^*\!\tilde F_{4B}
+\bar{\cal G}_{BC}(\bar z) \,{}^*\!F^C_{4}\right] \, ,\\
F^{i}_S &= \bar M z^{i}  
+ \frac{\ii}{2} {\cal M}^{iB}(z,\bar z) \left[\,{}^*\!\tilde F_{4B}
+  \bar{\cal G}_{BC}(\bar z) \,{}^*\!F^C_{4}\right] \, ,
\end{aligned}
\ee
where $z^i\equiv S^i|$ and $F^i_S\equiv -\frac14\cald^2 S^i|$ are the lowest and highest scalar components of the three-form multiplets $S^i$ ($i=1,\ldots,n$).

After having gauge-fixed the super-Weyl symmetry, the Lagrangian describing the coupling of the $S^i$ superfields to supergravity  takes the  form 
\be\label{OL}
\hat{\mathcal L}=-3\int\d^4\theta\,E\, e^{-\frac13 K(S,\bar S)} 
+ \left(\int\d^2\Theta\,2\cale\, \hat W(S)+\text{c.c.}\right)+{\mathcal L}_{\rm bd}\,.
\ee
Note that in  this Lagrangian the scalar auxiliary fields of the gravity and matter multiplets are defined by  \eqref{M+F} (ignoring fermions).

%%%%%%%%%%%%%%%%%%%%%%%%%%%%%%%%%%%%%%%%%%%%%%%%%%%%% 
%%%%%%%%%%%%%%%%%%%%%%%%%%%%%%%%%%%%%%%%%%%%%%%%%%%%% %%%%%%%%%%%%%%%%%%%%%%%%%%%%%%%%%%%%%%%%%%%%%%%%%%%%% 

\section{Application to type IIA  flux compactifications}
\label{sec:IIA}

As an application of the above dualization procedure, we will now consider an example of  type IIA  flux compactifications of string theory on a Calabi-Yau three-fold $CY_3$ in the presence of O6-planes. In particular, we will focus on the effective theory obtained by turning on RR fluxes in the internal $CY_3$ space. For simplicity, we will also set the internal NSNS flux $H_3$ to zero, so that the tadpole condition just requires that the O6 charge is cancelled by the presence of D6-planes, without involving the RR-fluxes. 

We will focus on the closed string scalar spectrum. The relevant terms in the effective $\caln=1$ supergravity for these kinds of compactifications can be found in \cite{Grimm:2004ua}. The closed string  moduli $v^i(x)$ and $b^i(x)$, $i=1,\ldots, h^{1,1}_-(CY_3)$,  are obtained by expanding the K\"ahler form $J$ and the NSNS two-form $B_2$ in a basis of orientifold-odd integral harmonic 2-forms $\omega_i\in H^2_-(X;\mathbb{Z})$ 
\be
J=v^i\omega_i\,,\qquad B_2=b^i\omega_i.
\ee
These moduli, together with their supersymmetric partners,  combine into $n\equiv h^{1,1}_-(CY_3)$ chiral superfields $\Phi^i$ with  lowest components 
\be \label{Moduli_Comp}
\Phi^i| = \varphi^i=v^i-\ii b^i \, . 
\ee
Furthermore, the complex structure, the dilaton, the internal RR three-form moduli and the associated supersymmetric partners combine into additional chiral superfields $T^q$, $q=1,\ldots, h^{2,1}(CY_3)+1$. The effective supergravity theory is characterised by the following K\"ahler potential \footnote{The formulas of \cite{Grimm:2004ua} are valid in the large volume and constant warping approximation, which then neglects the back-reaction of the fluxes and branes on the underlying Calabi-Yau geometry. The backreaction of fluxes and branes is expected to break the split structure of (\ref{K=K+K}).}  
\be\label{K=K+K}
\mathcal K(\Phi,\bar\Phi,T,\bar T)=K(\Phi,\bar\Phi)+ \hat K(T,\bar T). 
\ee
In the following we will not need the explicit form of $ \hat  K(T,\bar T)$, 
but we will just use the fact that it satisfies the condition 
\be\label{hatK}
 \hat  K^{\bar rq} \hat  K_q \hat  K_{\bar r}=4\,,
\ee
where, $ \hat  K_q\equiv\frac {\partial \hat  K}{\partial T^q}$, $ \hat  K_{q\bar r}\equiv\frac {\partial^2  \hat  K}{\partial T^q\partial \bar T^{\bar q}}$, \ldots, and  $ \hat  K^{\bar rq}$ is the inverse of the K\"ahler metric $ \hat  K_{q\bar r}$. 
Similarly, $K_{i}\equiv \frac {\partial K}{\partial \Phi^i}$, $ K_{i\bar\jmath}\equiv\frac {\partial^2  K}{\partial  \Phi^i\partial \bar\Phi^{\bar\jmath}}$, \ldots, and $K^{ \bar\jmath i}$ is the inverse of the K\"ahler metric $K_{ i\bar\jmath}$. 

The K\"ahler potential $K(\Phi,\bar\Phi)$ is given by
\be\label{K}
K(\Phi,\bar\Phi)=-\log\left[\frac1{3!}k_{ijk}(\Re\Phi^i)( \Re\Phi^j)(\Re\Phi^k)\right] \, , 
\ee
where $k_{ijk}$ are the intersection numbers
\be\label{Inters}
k_{ijk}=\int_{{\rm CY}_3} \omega_i\wedge\omega_j\wedge\omega_k.
\ee
Notice that $K(\Phi,\bar\Phi)$ depends only on the real combinations $\Phi^i+\bar\Phi^i$, so that we can make the identification $K_i\equiv K_{\bar\imath}$.  We will also use the fact that $K(\Phi,\bar\Phi)$ satisfies the no-scale condition 
\be\label{nos}
K^{\bar\jmath  i}K_{i}K_{\bar\jmath}-3=0.
\ee
The flux-induced superpotential is of the form introduced in \cite{Gukov:1999ya,Gukov:1999gr,Taylor:1999ii} 
and  depends only on the chiral  superfields $\Phi^i$ \footnote{In what follows we will tend to use notation close to that of \cite{Bielleman:2015ina,Carta:2016ynn}.}
\be\label{Wv}
W=e_0+\ii e_i\Phi^i-\frac 12 k_{ijk}m^i\Phi^j\Phi^k+\frac \ii6 m^0k_{ijk}\Phi^i\Phi^j\Phi^k,
\ee
where $e_0,e_i,m^i$ and $m^0$ represent the flux quanta of the internal RR fields. 

\subsection{Dualization to the three-form effective theory}

The effective theory described above has exactly the same structure as the theories considered in Section  \ref{sec:gensugra}, up to the explicit presence of a spectator sector given by the chiral fields $T^r$. In order to make this similarity manifest, we 
rewrite this theory in a super-Weyl invariant form by adding a super-Weyl compensator $Z$ and combining it with the chiral fields $\Phi^i$ into $n+1$ chiral superfields $\calz^A=(\calz^0,\calz^i)$ such that \be
\calz^0\equiv Z \, , 
\ee 
and 
\be\label{defZi}
\calz^i\equiv \ii Z \Phi^i \, , 
\ee  
which transform as in (\ref{weylZ}) under the super-Weyl transformations. Then it is easy to see that    
the superpotential (\ref{Wv}) gets transformed  into  (\ref{WWrel}) of the form 
\be\label{mF}
{\mathcal W}(\mathcal Z) =e_A\calz^A+\frac 1{2\calz^0} m^ik_{ijk}\calz^j\calz^k-\frac 1{6 (\calz^{0})^2} m^0k_{ijk}\calz^i\calz^j\calz^k\,.
\ee
This clearly satisfies the homogeneity condition  \eqref{OWhom} and can be written in the form (\ref{restrW}) with $\hat\calw(\calz)=0$ and 
$\calg_{AB}=\del_A\del_B\calg(\calz)$, where
\be\label{GFluxComp}
\calg(\calz)=\frac 1{6\calz^0} k_{ijk}\calz^i\calz^j\calz^k \,.
\ee

We are now in a position to apply the  duality transformation described in Section  \ref{sec:gensugra}. 
After dualization and gauge-fixing the super-Weyl symmetry by setting 
\be
Z=S^0=1 \, , 
\ee 
the final result is a Lagrangian of the form (\ref{OL}) with $\hat W=0$ and a K\"ahler potential which is modified by a contribution of the `spectators' $T^r$
\be\label{fluxdual}
\call=-3\int\d^4\theta\,E\, e^{-\frac13 K(S,\bar S)-\frac13 \hat K(T,\bar T)}+\call_{\rm bd}\,.
\ee 
Moreover, the superpotential has completely disappeared from the dual effective theory, since it is now encoded in the structure of the constrained superfields \eqref{mZA}. 

Notice that because of the definition (\ref{defZi}), after dualization and gauge-fixing we have $\Phi^i =-\ii\, S^i$ and we can identify the lowest components as follows
\be
 S^i|\equiv z^i=\ii \varphi^i.
\ee
In the following it will be also convenient to use 
\be
F^i\equiv - \ii  F^i_S 
\ee
instead of $F^i_S$, such that $-\frac14\cald^2\Phi^i=F^i$. 

Upon setting to zero the fermions, the independent bosonic components of these superfields are given by \eqref{z} and (\ref{M+F}). The latter take the following form in the case under consideration\footnote{To arrive at these relations we have used the specific form of the K\"ahler and superpotenional associated to the type IIA compactification in question given in Appendix \ref{C}. } 
 \be\label{summary1}
 \begin{aligned}
{\rm Re}\,M&=\frac 12\, {}^*\!\mathcal F^0_4\,,\\
{\rm Im}\,M&=
-2 e^{K}\,{}^*\!\tilde {\mathcal F}_{40}-\frac 12K_i\,{}^*\!\mathcal F^{i}_4\,,\\
{\rm Re} \,F^i
&=\frac 14 \,{}^*\!\mathcal F^0_4\,v^i-e^K(K^{ij}-2v^iv^j)\,{}^*\!\tilde {\mathcal F}_{4j}\,,\\
{\rm Im} \,F^i&=2 e^{K}\,{}^*\!\tilde {\mathcal F}_{40}\,v^i+\frac 12(\,{}^*\!{\mathcal F}_{4}^{{i}}+\,v^i\,K_j\,{}^*\!\mathcal F^{j}_4),
\end{aligned}
\ee 
where the four-forms $\calf_4^A$ and $\tilde\calf_{4A}$ are defined  in terms of the field-strengths $F_4^A=\d A^A_3$ and $\tilde F_{4A} =\d \tilde A_{3A}$ as follows 
\be\label{calfdef}
\begin{aligned}
\mathcal F^0_4&=-F^0_4\,, \qquad \mathcal F^i_4= -F^i_4+b^iF^0_4\,,\\
\tilde {\mathcal F}_{4i}&= \tilde F_{4i}+k_{ijk}b^jF_4^k-\frac 12 k_{ijk}b^jb^k F^0_4\,,\\
\tilde {\mathcal F}_{40}&= \tilde F_{40}+b^i\tilde F_{4i}+\frac 12 k_{ijk}b^ib^jF^k_4-\frac 16 k_{ijk}b^ib^jb^k F^0_4\,.
\end{aligned}
\ee
Note that the four-forms $\calf_4^A$ and $\tilde\calf_{4A}$ have exactly the same structure as the four-forms obtained in  \cite{Bielleman:2015ina,Carta:2016ynn} upon dimensional reduction of the type IIA RR field-strengths\footnote{The structure  of these field-strengths reflects the $B_2$-twisting of the ten-dimensional RR-fields in the  so called A-basis of the democratic formulation of \cite{Bergshoeff:2001pv}, which provides a duality-symmetric description of the type IIA supergravity theory, whose $E_{11}$ origin was revealed in \cite{West:2001as,Kleinschmidt:2003mf}.}, 
which is quite encouraging. 
To convince ourselves that this is not a mere coincidence,  in the following Section we will compute the on-shell values of \eqref{calfdef} by solving the equations of motion of $A^A_3$ and $\tilde A_{3A}$ which follow from the dual Lagrangian \eqref{fluxdual}.
As we will see, these on-shell values perfectly match those obtained by ten-to-four dimensional reduction  \cite{Bielleman:2015ina,Carta:2016ynn}.

The bosonic part of the dual Lagrangian \eqref{fluxdual} can be computed by  setting to zero fermionic component fields,  integrating over the Grassmann variables and integrating out the supergravity auxiliary vector field. Finally, one can go  to the Einstein frame by performing 
a Weyl re-scaling of the vielbeins, the dual four-form field strengths and the component fields in (\ref{summary1}) 
as follows 
\be
\label{eK}
\begin{split}
& e^a_m \to e^a_m e^{\frac {1}{6} (K+\hat K)}\,,\quad M  \to  M\,e^{-\frac 23 (K+\hat K)}\,, 
\\
& F^i  \to F^i\,e^{-\frac 23( K+\hat K)}\,, \quad F^q_T  \to F^q_T\,e^{-\frac 23( K+\hat K)}\, . 
\end{split}
\ee
The result is the following bosonic Lagrangian
\be\label{bosonicL}
e^{-1} {\cal L}_{\rm bos} =  -\frac{1}{2} R 
- K_{i\bar\jmath} (\varphi,\bar\varphi)\,\partial \varphi^i \partial \bar \varphi^{\bar\jmath} - \hat K_{q \bar r} (t,\bar t)\,\partial t^q \partial \bar t^{\bar r} + e^{-1}\,{\cal L}_{\text{3-form}} \, , 
\ee
in which  $t^q\equiv T^q|$ and ${\cal L}_{\text{3-form}}$ contains  the three-form sector encoded in $M$ and $F^i$ as in (\ref{summary1}) and the auxiliary fields $F^q_T$ of the $T^q$ multiplets 
\be\label{S+MLW2} 
\begin{aligned}
e^{-1} {{\cal L}}_{\text{3-form}} =&\,  e^{- \calk}K_{i\bar\jmath} F^i \bar F^{\bar\jmath}+ e^{- \calk} \hat K_{q \bar r} F^q_T \bar F_T^{\bar r} \\
&- \frac{1}{3} e^{-\calk}\left(M+K_{\bar\imath} \bar F^{\bar\imath}+\hat K_{\bar q} \bar F^{\bar q}_T\right) 
\left( \bar M+K_{i}F^{ i}+\hat K_{q}F^{q}_T \right)+\call_{\rm bd}\,.
\end{aligned}
\ee
where we recall that $\calk\equiv K+\hat K$.
With the help of  the no-scale condition \eqref{hatK}, we can easily integrate out the auxiliary fields $F^q_T$ by solving their equations of motion, whose solution is  
\be
\hat K_{q \bar r}\bar F^{\bar r}_T=-\left(M+K_{\bar\imath} \bar F^{\bar\imath}\right)\hat K_q\,.
\ee
Substituting it back into the Lagrangian \eqref{S+MLW2} and using \eqref{summary1} we obtain the following Lagrangian for the gauge three-forms \be\label{Laux2new}
\begin{split}
e^{\hat K} e^{-1} {\cal L}_{\text{3-form}} 
= & \, \frac{e^{-K}}{16} \left( {\,}^* {\cal F}_4^0 \right)^2 
+ e^K K^{ij} {\,}^* \tilde{\cal F}_{4i} {\,}^* \tilde{\cal F}_{4j} 
\\ 
&+ \frac{e^{-K}}{4} K_{ij} {\,}^* {\cal F}_4^i {\,}^* {\cal F}_4^j 
+ 4 e^K \left( {\,}^* \tilde{\cal F}_{40} \right)^2 +\call_{\rm bd}\, , 
\end{split}
\ee 
with the boundary term 
\be\label{NLboundary1}
\begin{aligned}
{\mathcal L}_{\rm bd}=&\, -2\partial_m\left[e\,\tilde A^m_0\, 4e^{K-\hat K}\,{}^*\!\tilde {\mathcal F}_{40}+e\,\tilde A^m_i\,e^{K-\hat K}\left(K^{ij}\,{}^*\!\tilde {\mathcal F}_{4j}+4b^i\,{}^*\!\tilde {\mathcal F}_{40}\right)\right]\\
&+2\partial_m\Big[e\, A^{mi}\,e^{-\hat K}\Big(\frac14 {e^{- K}} \,K_{ij}\,{}^*\!{\mathcal F}_{4}^{{j}}-
k_{ijk}b^j e^KK^{kl}\,{}^*\!\tilde {\mathcal F}_{4l}-2 k_{ijk}b^jb^k \,{}^*\!\tilde {\mathcal F}_{40}\Big)\Big]\\
&+2\partial_m\Big[e\, A^{m0}\,e^{-\hat K}\Big(\frac1{16} e^{- K}\,{}^*\!\mathcal F^0_4+\frac{e^K}2k_{ijk}b^jb^kK^{il}\,{}^*\!\tilde {\mathcal F}_{4l}-\frac{e^{-K}}{4}b^iK_{ij}\,{}^*\!{\mathcal F}_{4}^{{j}}\\
&\quad~~~~~~~~~~~~~~~~~~~~~~~~~~~~~~~~~~~~~~~~~~~~~~~~~~~~~~~+ \frac {2}3\,k_{ijk}b^ib^jb^k\,e^{K}\,{}^*\!\tilde {\mathcal F}_{40}\Big)\Big],
\end{aligned}
\ee
where we recall that
$A^{A}_m\equiv({}^*A^A_3)_m$ and  $\tilde A_{Am}\equiv({}^*\tilde A_{3A})_m$. 
This boundary term is directly extracted by writing the superspace Lagrangian \eqref{NLboundary} in components. 

The  Lagrangian (\ref{bosonicL})-(\ref{NLboundary1}) provides a non-trivial example of the effect of the non-linear dualization procedure put forward in this paper.  We explicitly see that it does not depend on the constants $e_A$ and $m^A$ appearing in the original Lagrangian and does not contain any potential for the scalar fields. Rather, as we will discuss in the next Section, it is  generated dynamically by the gauge three-forms  $A^A_3$ and $\tilde A_{3A}$.

\subsection{Back to the original theory}

Let us show how the bosonic Lagrangian of the original theory is reproduced from the bosonic Lagrangian (\ref{bosonicL}). This is done by integrating out the gauge three-forms  $A^A_3$ and $\tilde A_{3A}$ which enter  $\calf_4^A$ and $\tilde\calf_{4A}$ as defined in (\ref{calfdef}). Indeed, the integration of the equations of motion which follow from  (\ref{Laux2new}) produces the following expressions involving  $2n+2$ integration constants which, for obvious reasons,  we call $e_A$  and $m^A$  
\be\label{F_40m}
\begin{aligned}
-4e^{-\hat K}e^K\,{}^*\!\tilde {\mathcal F}_{40}=m^0\,,\\
-e^{-\hat K}e^K\,K^{ij}\,{}^*\!\tilde {\mathcal F}_{4j}=m^i-m^0 b^i\equiv p^i\,,\\
-\frac 14 e^{-( K+\hat K)}\,K_{ij}\,{}^*\!{\mathcal F}_{4}^{{j}}=e_i + k_{ijk}b^jm^k-\frac 12 k_{ijk}b^jb^k\,m^0\,\equiv \rho_i\,,\\
-\frac 1{16}e^{-(K+\hat K)}\,{}^*\!\mathcal F^0_4=e_0+b^i e_i+\frac 12k_{ijk}b^ib^jm^k-\frac 16 k_{ijk}b^ib^jb^km^0\equiv \rho_0\,.
\end{aligned}
\ee
These are exactly (modulo some conventions) the on-shell values of the four-forms obtained in \cite{Bielleman:2015ina,Carta:2016ynn} by dimensionally reducing the ten-dimensional Hodge duality relations between the type IIA RR field-strengths.

Substituting  (\ref{F_40m}) back into the bosonic Lagrangian (\ref{Laux2new}) and \eqref{NLboundary1}, one obtains the scalar potential of the original theory which coincides with the well-known form of the type IIA RR flux potential \cite{Louis:2002ny,Grimm:2004ua}
\be
\begin{aligned}
V &=-e^{-1}\, \call_{\text{3-form}}|_{\text{on-shell}} =\\
&= e^{\hat{K}} \Big[ 16 e^K\, \rho_0^2 + 4 e^{K}\,K^{ij}\rho_i \rho_j + e^{-K}\,K_{ij}p^ip^j+ \frac14{(m^0)^2} e^{-K} \Big].\label{Veff}
\end{aligned}
\ee
Note that upon this substitution the term \eqref{NLboundary1} is no more a total derivative. Without the contribution of this term, the effective scalar potential would have a wrong (negative) sign. This is why we have kept track of the boundary terms in our construction all the time.

Note also that, if we substitute the on-shell values \eqref{F_40m} of the four-forms $\calf_4^A$ and $\tilde\calf_{4A}$ into the boundary Lagrangian \eqref{NLboundary1}, while still keeping the potentials $A^{mA}$ and $\tilde A^{m}_{A}$ off-shell, upon some algebra we get
\be\label{NLboundary2}
\begin{aligned}
\hat{\mathcal L}_{\rm bd}&=2e\left(\rho_0\,{}^*\!\mathcal F^0_4+\rho_i\,{}^*\!{\mathcal F}_{4}^{{i}}+p^i\,{}^*\!\tilde {\mathcal F}_{4i}+m^0\,{}^*\!\tilde {\mathcal F}_{40}\right)\\
&= 2\partial_m\,\left(e\,\left(m^A\, \tilde A^m_A-e_A\,A^{Am}\right)\right)\,.
\end{aligned}
\ee
This boundary term is the same as the linear term of the effective Lagrangian obtained in \cite{Bielleman:2015ina} by the dimensional reduction of the democratic type IIA  pseudo-action of  \cite{Bergshoeff:2001pv}. It is a total derivative because of the use of the ten-dimensional Hodge duality relations between the lower- and higher-form RR field strengths, which, as we have already mentioned, are equivalent to the on-shell expressions \eqref{F_40m} for the four-forms (see \cite{Bielleman:2015ina} for details).
To perform the off-shell dimensional reduction one could use the fully-fledged duality-symmetric action of type IIA supergravity constructed in \cite{Bandos:2003et}. In this way, in principle, one should get  the four-dimensional Lagrangian \eqref{Laux2new} with the boundary term \eqref{NLboundary1}, which produces the constants $e_A$ and $m^A$ after one integrates out the 3-forms.

\section{Conclusion}

In this paper we have shown how to construct globally and locally supersymmetric models with gauge three-forms, by dualising more conventional theories with standard chiral multiplets and  a superpotential of the form (\ref{sup0}). In the dualization process, the coupling constants $e_A,m^B$ are promoted to (appropriate combinations of) expectation values of the field-strengths $F^A_4=\d A^A_3$, $\tilde F_{A4}=\d \tilde A_{A3}$ associated with the three-form gauge fields $A^A_3$, $\tilde A_{A3}$. The dual theory is manifestly supersymmetric and is constructed in terms of three-form multiplets which contain a complex scalar and one or two gauge three-forms as bosonic components, the latter replacing scalar auxiliary fields of the conventional chiral multiplets.

As an application, we applied our duality procedure to the four-dimensional effective theory describing  the closed string sector of type IIA orientifold compactifications on Calabi-Yau three-folds with RR fluxes. In particular, we discussed the explicit form of the bosonic action for the scalar and three-form fields. By solving the equations of motion for  the three-form fields we found the same on-shell values of their  field-strengths as those  obtained by direct dimensional reduction \cite{Bielleman:2015ina} and the correct potential for the scalar fields \cite{Grimm:2004ua}.  

Even though our approach is quite general, the application to more general string compactifications requires further work. First of all, in the type IIA models considered in Section \ref{sec:IIA} the tadpole condition does not directly concern the internal fluxes that are involved in the dualization.   
In more general IIA compactifications, for instance with a non-trivial $H_3$-flux, the tadpole condition would become relevant for the dualization procedure. The same is true for type IIB orientifold compactifications, which have flux induced superpotential \cite{Gukov:1999ya,Gukov:1999gr,Taylor:1999ii,Giddings:2001yu,Grimm:2004uq}  compatible with our general framework too. Also in these cases a non-trivial tadpole condition should be appropriately taken into account. 

Another aspect that deserves further study is the inclusion of the open-string sector in the effective theory, which may be naturally incorporated in a three-form formulation  \cite{Grimm:2011dx,Kerstan:2011dy,Carta:2016ynn}. It would be interesting to revisit this point in the manifestly supersymmetric framework provided in the present paper. Related questions concern its applications to M-theory and  F-theory compactifications, which can be considered as strong coupling limits of type IIA and IIB compactifications with backreacting branes, see for instance \cite{Acharya:2004qe,Denef:2008wq} for reviews. 

In four dimensions, gauge three-forms couple to membranes that appear as domain-walls generating jumps of the value of the corresponding field-strength, as e.g.\ in \cite {Brown:1987dd}. 
In the context of string/M-theory compactifications, these membranes correspond to higher-dimensional branes wrapping various cycles in the internal space and are crucial for the mechanisms of dynamical relaxation  of the cosmological constant discussed, for instance, in \cite{Bousso:2000xa,Feng:2000if}.
Our  formulation should be the starting point for revisiting these aspects at the level of a four-dimensional  effective theory with manifest linearly-realized supersymmetry,
 generalizing the results of \cite{Ovrut:1997ur,Bandos:2010yy,Bandos:2011fw,Bandos:2012gz}. 
Furthermore, in this same context and in the presence of spontaneously broken supersymmetry, our formulation should be related, at low energies, to models with non-linearly realized local supersymmetry as the ones introduced in  \cite{Farakos:2016hly}. It would be interesting to elucidate this relation. More generically, it would be worth using this general framework to construct and study supersymmetric extensions of various models based on gauge three-forms, as for instance those discussed in \cite{Duff:1980qv,Hawking:1984hk,Brown:1987dd,Brown:1988kg,Duff:1989ah,Duncan:1989ug,Wu:2007ht,Dvali:2005an,Dvali:2004tma,Dvali:2005zk,Dvali:2013cpa,Kaloper:2008fb,Kaloper:2011jz,Marchesano:2014mla,Bielleman:2015ina,Dudas:2014pva}.

\subsection*{Acknowledgements}
We thank Igor Bandos, Massimo Bianchi, Sergei Kuzenko and Irene Valenzuela for useful discussions and comments. 
Work of F.F., S.L.\ and L.M.\ was partially supported by the Padua University Project CPDA144437.
Work of D.S. was supported in part by the Australian Research Council project No. DP160103633 and by the Russian Science Foundation grant 14-42-00047 in association with Lebedev Physical Institute.

\begin{appendix}
\section{Component structure of ${\cal N}=1$ superfields}\label{B}

In this appendix we collect some useful formulas on the component structure of the multiplets considered in the present paper. We mostly follow notation and conventions of \cite{Wess:1992cp}. 

The chiral multiplet $\Phi$ is defined by the condition 
\be
\bar{D}_{\dot\alpha} \Phi = 0.
\label{eq:ChiralA}
\ee
Its component expansion is 
\be
\Phi = \varphi + \sqrt{2} \theta \psi + \theta^2 F + \ii \theta \sigma^m \bar\theta \partial_m \varphi -\frac{\ii}{\sqrt{2}} \theta^2 \partial_m \psi \sigma^m \bar\theta + \frac 14 \theta^2\bar\theta^2 \Box \varphi,
\label{ChiralB}
\ee
where $\varphi$ and $F$ are complex scalar fields and $\psi$ is a Weyl spinor. 
The independent bosonic components of $\Phi$ are defined as follows
\be\label{ChiralComp}
\begin{aligned}
\Phi | &= \varphi , \\
-\frac14 D^2 \Phi| &= F,
\end{aligned}
\ee
where the vertical line means that the quantity is evaluated at $\theta=\bar\theta=0$.
The real scalar multiplet $U$ has the following component structure 
\be
\begin{split}
U =& \, u + \ii \theta \chi  - \ii \bar\theta \bar\chi + {\ii} \theta^2 \bar\varphi -  {\ii} \bar\theta^2 {\varphi} 
+2 \theta \sigma^m \bar\theta A_m  
\\
    & +\ii\theta^2\bar{\theta} \left( \bar{\lambda} 
    +\frac{\ii}{2}\bar{\sigma}^m\partial_m \chi\right)-\ii\bar{\theta}^2 \theta \left( \lambda+\frac{\ii}{2}{\sigma}^m\partial_m \bar{\chi}\right)- \theta^2\bar\theta^2 \left(D+\frac14\Box u \right) \, , 
\label{VectorB}
\end{split}
\ee
where $u$ and $D$ are real scalar fields, $\varphi$ is a complex scalar field, $A_m$ is a real vector field and $\chi$ and $\lambda$ are Weyl spinors. 
The independent bosonic components of $U$ are defined as follows
\be
\begin{aligned}
U | &= u ,\\
- \frac 18 \bar{\sigma}^{\dot\alpha \alpha}_m [D_\alpha, \bar{D}_{\dot\alpha}] U| &= A_m,\\ 
\frac{\ii}{4} D^2 U| &= \bar\varphi,\\
\frac{1}{16} D^2 \bar{D}^2 U| &= -D + \ii \partial^m A_m  \, .
\end{aligned}
\ee
The real linear multiplet $L$ is a real multiplet which, in addition, satisfies the condition
\be
D^2 L = 0, \qquad \bar{D}^2 L = 0 .
\label{RealLMdef}
\ee
The component expansion of $L$ has the form
\be
\begin{split}
L = &l + \ii \theta \eta -\ii \bar\theta \bar\eta + \frac13 \theta \sigma_m \bar\theta \varepsilon^{mnpq} \partial_{[n}\Lambda_{pq]} 
\\
&+ \frac12 \theta^2\bar\theta \bar\sigma^m \partial_m \eta - \frac12 \theta \bar\theta^2 \sigma^m \partial_m \bar\eta -\frac14 \theta^2\bar\theta^2 \Box l \, , 
\end{split}
\label{RealLM}
\ee
where $l$ is a real scalar,  $\Lambda_{mn}$ is a rank 2 antisimmetric tensor  and $\eta$ is a Weyl spinor. \\
The bosonic components of $L$ are defined through the projections 
\be
\begin{aligned}
L | &= l, \\
\frac12\bar{\sigma}^{m\,\dot\alpha \alpha}\left[D_\alpha,\bar{D}_{\dot{\alpha}}\right] L| &=  - \frac23 \varepsilon^{mnpq} \partial_{[n}\Lambda_{pq]} .
\end{aligned}
\ee
The complex linear multiplet $\Sigma$ satisfies the condition
\be
\bar{D}^2 \Sigma = 0 .
\label{ComplexLMdef}
\ee
Its $\theta$-expansion is 
\be
\begin{split}
\Sigma = &\sigma + \theta \psi + \sqrt{2} \bar\theta \bar\rho -\frac12 \theta \sigma_m \bar\theta C^{m} + \theta^2 \bar s+ \theta^2\bar\theta \bar\xi 
\\
&-\frac{\ii}{\sqrt{2}} \bar\theta^2 \theta \sigma^m \partial_m \bar\rho + \theta^2\bar\theta^2 \left(\frac{\ii}{4} \partial_m C^{m} -\frac14 \Box \sigma \right).
\end{split}
\label{eq:ComplexLM}
\ee
Here $\sigma$ and $\bar s$ are complex scalars, $\rho$, $\psi$ and $\xi$ are Weyl spinors and $C^{m}$ is a complex vector which is Hodge dual to the three-form 
\be
C^{m} = \frac{1}{3!} \varepsilon^{mnpq} C_{npq}.
\label{ComplexLMvector}
\ee
The bosonic components of $\Sigma$ are defined  by the projections
\be\label{Sigmaexp}
\begin{aligned}
\Sigma | &= \sigma ,\\
\frac12 \bar{\sigma}^{m\,\dot\alpha \alpha}
 \left[D_\alpha,\bar{D}_{\dot{\alpha}}\right] \Sigma| &= C^{m},\\
-\frac14 D^2 \Sigma| &= \bar s,\\
\frac{1}{16} D^2 \bar{D}^2 \Sigma| &= 0,\\
\frac{1}{16} \bar{D}^2 D^2 \Sigma| &= \frac{\ii}{2} \partial_m C^m.
\end{aligned}
\ee

\section{Note on three-forms, scalar potentials and boundary terms}\label{A}
In this Appendix we illustrate the dualization procedure with two simple examples: first we will consider a purely bosonic Lagrangian of a single gauge three-form and then we will examine the case of a Lagrangian with a single complex linear multiplet.

Let us consider a real three-form with couplings described by the Lagrangian 
\begin{equation}
\label{LC1}
{\cal L} = K''(\varphi) \left( \frac 1{3!}\partial_m \varepsilon^{mnpq} A_{npq} \right)^2 + W'(\varphi) \left(\frac 1{3!} \partial_m \varepsilon^{mnpq} A_{npq} \right) \, ,  
\end{equation} 
where $K''(\varphi)$ and $W'(\varphi)$ are real functions of the scalar fields $\varphi$, denoted in this way to be reminiscent of the structure of supersymmetric chiral field models. 
To further simplify the formulas, let us replace $A_{npq}$ with its Hodge-dual vector field
\begin{equation}\label{Bound_Dual}
A^m = \frac{1}{3!} \varepsilon^{mnpq} A_{npq} \, , 
\end{equation}
so that \eqref{LC1} becomes 
\begin{equation}
\label{LC1B}
{\cal L} = K''(\varphi) \left( \partial_m A^m \right)^2 + W'(\varphi) \left(\partial_m A^m \right) \, .  
\end{equation}
Note that the gauge invariance of the three form becomes an invariance of the action under 
the transformation of the one-form   $A_1 \to A_1 + {\,}^* \! \d \Lambda_2 $, 
where $A_1= A_m \d x^m$.

We wish to integrate out $A^m$ to find the contribution to the scalar potential. 
To perform a consistent variation of the action with respect to the three-form, 
one should introduce an appropriate boundary term of the form \cite{Brown:1987dd,Duff:1989ah} 
\be \label{LCBound}
{\cal L}_{\textrm{bd}} = - \partial_m \left( (W' + 2 K'' \, \partial_n A^n ) \, A^m  \right) \, .  
\ee
Then the equations of motion for the three-form (which are unaffected by the boundary terms) 
give 
\be \label{LConshell}
\partial_m A^m = - \frac{W' + r}{2 K''} \, ,
\ee
where $r$ is a real integration constant. Substituting \eqref{LConshell} into \eqref{LC1B}+\eqref{LCBound} we get
the following Lagrangian which provides the potential for the scalar fields $\varphi$ 
\begin{equation}
\label{LC4}
{\cal L} = - \frac{( r+ W' )^2}{4 K''} \, . 
\end{equation} 
There is an alternative way to integrate out the three-form 
without the need to introduce the boundary terms. 
We can rewrite \eqref{LC1B} by using a Lagrange multiplier scalar $\alpha$ and an auxiliary field $F$
\begin{equation}
\label{LC2}
{\cal L} = K'' F^2 
+ W' F  
+ \alpha F + A^m\partial_m \alpha  \, .   
\end{equation}
By varying $\alpha$ in \eqref{LC2} with the boundary condition $\delta \alpha|_{\textrm{bd}}=0$ we get $F=\partial_m A^m$ and then back \eqref{LC1B}. 
Now $A^m$ is a Lagrange multiplier and we can consistently integrate it out 
without the need of additional boundary terms thus getting
\begin{equation}
\alpha = r \, , 
\end{equation}
where $r$ is a real constant related to the on-shell value of $F_4=dA_3$. 
Now we have 
\begin{equation}
\label{LC3}
{\cal L}= K'' F^2 + ( r + W' ) F \, ,   
\end{equation}
and once we integrate out the scalar $F$ we find \eqref{LC4} 
which produces a positive definite contribution to the scalar potential (if $K''>0$). 

On the other hand, this dualization procedure provides a systematic way to get the boundary term \eqref{LCBound} which is necessary to make the variation of the Lagrangian \eqref{LC1} consistent. 
To do this we should reverse the dualization procedure starting from the Lagrangian \eqref{LC2}. 
The variation of the Lagrangian \eqref{LC2} with respect to the auxiliary field gives
\be
\begin{split}
\delta_\alpha \call &= \delta \alpha \left( F - \partial_m A^m \right) + \partial_m (A^m \delta \alpha ) \, , 
\\
\delta_F \call &=  \left(2 K'' F + \alpha + W' \right) \delta F\,.
\end{split}
\ee
Imposing the boundary conditions
\begin{equation}\label{Bound_Cond}
\delta \alpha |_{\mathrm{bd}} = 0, \qquad \delta F |_{\mathrm{bd}} = 0 \, , 
\end{equation}
and setting the variations to zero
we get
\be
\alpha = - 2K'' F - W' \,,\qquad
F = \partial_m A^m \label{Bound_Aux_2}.
\ee
Plugging  \eqref{Bound_Aux_2} back into the Lagrangian \eqref{LC2}, we get 
\begin{equation}\label{Bound_Action_1B}
\mathcal L =  K'' (\partial_m A^m)^2  + W' \partial_m A^m  - \partial_m \left( A^m (W'+ 2K''\partial_n A^n) \right) \, , 
\end{equation}
which reproduces the boundary term \eqref{LCBound}.

Let us now consider an example which shows how a Lagrangian of the form \eqref{LC1B} can be obtained by the direct computation of the bosonic components of a superspace Lagrangian of the form discussed in Section \ref{rigid}. Let us consider the following Lagrangian for a single chiral multiplet $\Phi$ (whose component structure was given in \eqref{ChiralComp})
\begin{equation}\label{Bound_Action_Susy_A}
\call= \int \d^4\theta\, \Phi \bar{\Phi} + \left(\int \d^2\theta \left(c\,\Phi+W(\Phi)\right) + c.c. \right) \, , 
\end{equation}
with $c$ being a complex constant. 
In order to make the auxiliary field $F$ of $\Phi$ dynamical, we  promote the complex constant $c$ to a chiral superfield $X$ and add a new term which contains the complex linear superfield $\Sigma$ 
\begin{equation}\label{Bound_Action_Susy_B}
\begin{split}
\call=& \int \d^4\theta\, \Phi \bar{\Phi} + \left(\int \d^2\theta \left(X\,\Phi+W(\Phi)\right) + c.c. \right) 
\\ 
&+\left[ \int \d^2 \theta \left(-\frac14 \bar{D}^2 \right) \left(\bar{X} \Sigma \right) + c.c \right] \, . 
\end{split}
\end{equation} 
Using the expansions of the superfields in component fields given in Appendix \ref{B} and focusing on the bosonic components only, we get from \eqref{Bound_Action_Susy_B} the following part of the component Lagrangian which contains the auxiliary fields $F$ and $\bar F$
\begin{equation}\label{Bound_Action_Susy_C}
\begin{split}
\call_F=& F\bar{F} + \left( W'F + \alpha F +\frac{\ii}{2}\,C^m\, \del_m \alpha + c.c.\right) \, , 
\end{split}
\end{equation}
where $\alpha= X|$ and, as usual, the vector field $C^m$ is dual of a three form.
This is a complexified version of the Lagrangian \eqref{LC2}.

To obtain the dual Lagrangian for the fields $C^m$ we vary \eqref{Bound_Action_Susy_C} with respect to $\alpha $ and $F$, and get the equations of motion
\be
\label{Bound_Action_Susy_D}
\begin{aligned}
F &= \frac{\ii}{2}\, \partial_m C^m,\\
\alpha &= - \bar{F} -W'\,.
\end{aligned}
\ee
Plugging them back into \eqref{Bound_Action_Susy_C}, we get
\begin{equation}\label{Bound_Action_Susy_E}
\call_F= \frac14 \left(\partial_n C^n\right)\left(\partial_m \bar{C}^m\right) + \left( \frac{\ii}{2}\, W'\partial_n C^n + c.c.\right) + \call_{\rm{bd}} \, , 
\end{equation}
with the boundary term Lagrangian having the required form 
\begin{equation}\label{Bound_Action_Susy_E1}
\call_{\rm{bd}}= \frac{\ii}{2}\, \partial_m \left(\left(\frac{\ii}{2}\, \partial_n \bar{C}^n - W'\right) C^m \right)  + c.c.
\end{equation}

\section{Properties of the K\"ahler potential and superpotential of type IIA compactifications with RR fluxes}\label{C}

Here we give some useful expressions that we used for the analysis of the effective four-dimensional theory associated with the example of type IIA flux compactification in Section \ref{sec:IIA}.

The $K$ part \eqref{K} of the K\"ahler potential \eqref{K=K+K} of the model under consideration is
\be\label{RelKk}
K=-{\rm log}\,8k,
\ee
where
\be \label{Inters_b}
k = \frac{1}{3!} k_{ijk} v^i v^j v^k\,, \qquad  k_{ij} \equiv k_{ijk} v^k, \qquad k_i \equiv k_{ijk} v^j v^k
\ee
and $k_{ijk}$ is the triple intersection number of the $CY_3$ manifold.

Defining  $K_{i}\equiv \frac {\partial K}{\partial \varphi^i}$ and $K_{ij}\equiv\frac {\partial^2  K}{\partial  \varphi^i\partial \bar\varphi^{j}}$, we have
\be
\begin{aligned}
&K_i =-\frac {k_i}{4k}=-2k_i{e^K},\\
&K_{ij}=-\frac 1{4k}\left(k_{ij}-\frac{k_ik_j}{4k}\right),\\
&K^{ij} \equiv \left(K_{ij}\right)^{-1} = -4k\left(k^{ij}-\frac{v^iv^j}{2k}\right),
\end{aligned}
\ee
and
\be
K^{ij}K_j=-2v^i,\qquad K_{ij}v^j=-\frac 12 K_i,\qquad K_iv^i =-\frac 32\,.
\ee
From \eqref{GFluxComp}, upon gauge-fixing $\calz_0 =1$, we get the following components of the imaginary and the real parts of the holomorphic matrix $\calg_{AB}$ \eqref{splitgAB}
\be\label{ImG}
\calm_{00}=-2k+k_{ij}b^ib^j,\qquad \calm_{0i}=-k_{ij}b^j=- \calm_{ij}b^j\,,\qquad \calm_{ij}= k_{ij}\, , 
\ee
\be\label{ReG}
\caln_{00}=\frac 13 k_{ijk}b^ib^jb^k-k_{i}b^i,\qquad \caln_{0i}=\frac 12 \left(k_i-k_{ijk}b^jb^k\right),\qquad
\caln_{ij}=k_{ijk}b^k\,.
\ee
The inverse matrix $\calm^{AB}$ has the following components 
\be\label{M-1}
\mathcal M^{00}=-\frac 1{2k}\,,\qquad  \mathcal M^{0i}=-\frac 1{2k} b^i\,,\qquad 
\mathcal M^{ij}=k^{ij}-\frac 1{2k} b^ib^j\,.
\ee

\end{appendix}

%\maketitle  IS IGNORED %%%%%%%%%%%

%\listoftables       % ONLY IN DRAFT MODE
%\listoffigures      % ONLY IN DRAFT MODE

%%%%%%%%%%%%%%%%%%%%%%%%%%%%%%
%%%%%%%%%%%%%%%%%%%%%%%%%%%%%%

%\if{}
%\bibliographystyle{abe}
%\bibliography{references}{}
%\fi

%%%%%%%%%%%%%%%%%%%%%%%%%%%%%%
%%%%%%%%%%%%%%%%%%%%%%%%%%%%%%

\providecommand{\href}[2]{#2}\begingroup\raggedright\endgroup

%%%%%%%%%%%%%%%%%%%%%%%%%%%%%%
%%%%%%%%%%%%%%%%%%%%%%%%%%%%%%

\end{document}